\documentclass[12pt,preprint]{aastex}

\usepackage{amssymb}
\usepackage{amsmath}

\shorttitle{PopIII stars formation} 
\shortauthors{Trenti \& Stiavelli}

\begin{document}

\newcommand{\msun}{\mathrm{M_{\sun}}}
\newcommand{\be}{\begin{equation}}
\newcommand{\ee}{\end{equation}}


\title{The formation rates of Population III stars and chemical
enrichment of halos during the Reionization Era}


\author{Michele Trenti}
\affil{University of Colorado, Center for Astrophysics and Space Astronomy, 389-UCB, Boulder, CO 80309 USA; Space Telescope Science Institute, 3700 San Martin Drive Baltimore MD 21218 USA}
\email{trenti@colorado.edu}
\author{Massimo Stiavelli}
\affil{Space Telescope Science Institute, 3700 San Martin Drive Baltimore MD 21218 USA}


\begin{abstract}

The First Stars in the Universe form out of pristine primordial gas
clouds that have been radiatively cooled to a few hundreds of degrees
Kelvin either via molecular or atomic (Lyman-$\alpha$) hydrogen
lines. This primordial mode of star formation is eventually quenched
once radiative and/or chemical (metal enrichment) feedbacks mark the
transition to Population II stars. In this paper we present a model
for the formation rate of Population III stars based on
Press-Schechter modeling coupled with analytical recipes for gas
cooling and radiative feedback. Our model also includes a novel
treatment for metal pollution based on self-enrichment due to a
previous episode of Population III star formation in progenitor
halos. With this model we derive the star formation history of
Population III stars, their contribution to the re-ionization of the
Universe and the time of the transition from Population III star
formation in minihalos ($M \approx 10^{6} \mathrm{M_{\sun}}$, cooled
via molecular hydrogen) to that in more massive halos ($M \gtrsim 2
\times 10^{7} \mathrm{M_{\sun}}$, where atomic hydrogen cooling is
also possible). We consider a grid of models highlighting the impact
of varying the values for the free parameters used, such as star
formation and feedback efficiency. The most critical factor is the
assumption that only one Population III star is formed in a halo. In
this scenario, metal free stars contribute only to a minor fraction of
the total number of photons required to re-ionize the universe. In
addition, metal free star formation is primarily located in minihalos
and chemically enriched halos become the dominant locus of star
formation very early in the life of the Universe --- at redshift $z
\approx 25$ --- even assuming a modest fraction ($0.5$\%) of enriched
gas converted in stars. If instead multiple metal free stars are
allowed to form out of a single halo, then there is an overall boost
of Population III star formation, with a consequent significant
contribution to the re-ionizing radiation budget. In addition, the
bulk of metal free stars are produced in halos with $M \gtrsim 2
\times 10^{7} \mathrm{M_{\sun}}$.
\end{abstract}


\section{Introduction}

Population III stars are considered to be the first luminous objects
formed during the Dark Ages of the Universe, when the hydrogen is in a
neutral state (e.g. see \citealt{bromm04}). The first generation of
stars, formed out of pristine primordial gas, had a top-heavy initial
mass function, with a typical mass scale of order of $\approx 100
\mathrm{M_{\sun}}$ and most probably just one star per halo
\citep[e.g. see][]{abel02,oshea07}. These stars start forming after
about $30-40$ million years from the big-bang at redshift $z \approx
55-65$ (\citealt{nao06,ts07a}; see also \citealt{gao05}) and, given
their high mass, they live only a few million years ending with either
a pair instability supernova phase or a direct collapse to a black
hole \citep{heger03}.

Population III stars thus initiate the chemical enrichment of the
Universe and open the way to more normal modes of star formation,
namely Population II \citep[e.g. see][]{ost96,furl03}. In fact, the
metals released into the IGM after a pair instability supernova
explosion can travel outside the parent dark matter halo that hosts
the Population III star. Calculations by \citet{brom01} found that a
region containing up to about $10^{8} \mathrm{M_{\sun}}$ can be
enriched to a critical metallicity $Z_{crit} \gtrsim 10^{-4} Z_{\sun}$
by the most massive pair instability supernovae. More typical
explosions may instead enrich significantly less gas ($\approx 10^{6}
\mathrm{M_{\sun}}$) although at a correspondingly higher metallicity
(see \citealt{bromm03,kita05,greif07,whalen08}). Even in the latter
case, a halo of mass $\lesssim 10^{8} \mathrm{M_{\sun}}$ that had one
of its progenitors hosting a Pair Instability Supernova is still
likely to be enriched to an average metallicity of $\gtrsim 10^{-4}
Z_{\sun}$ thanks to violent relaxation mixing \citep{lyn67} during its
hierarchical build-up.

Population III stars are also the sources that start to re-ionize the
Universe, creating ionized islands within the neutral hydrogen
inter-stellar and inter-galactic medium. Ionizing photons are emitted
with an enhanced efficiency
compared to Population II stars 
due to the high effective temperatures of massive metal-free stars
\citep{tum00,scha02}, and these sources could be responsible for a
significant fraction of the Thompson optical depth to reionization
deriving from $z>7$ \citep{shu08}. Another hint suggesting that
Population III stars contribute significantly to the re-ionization of
hydrogen can also be inferred by the rapid evolution of the galaxy
luminosity function at $z>6$, which implies that observed galaxies
alone do not seem capable of re-ionizing the Universe (e.g. see
\citealt{oesch08,bou06}).

Two main modes of Population III star formation have been proposed:
either in minihalos with virial temperature of $T_{vir} \approx 10^3
K$, where the gas is cooled via molecular hydrogen ($H_2$), or in more
massive, rarer, halos with $T_{vir} \approx 10^4 K$, where cooling
through atomic hydrogen (Lyman-$\alpha$) lines becomes possible
\citep[e.g. see][]{bromm04}. $H_2$ is formed during the initial
collapse of the gas within the minihalo, but it is sensitive to
photo-dissociating radiation in the Lyman Werner band ($[11.18:13.60]
\mathrm{eV}$). Thus, in presence of a sufficiently high LW background,
its formation rate may be lower than the dissociation rate with a
resulting suppression of Population III star formation in minihalos
\citep{hai97,hai00,cia00,glo01,macha01,oshea08}. Interestingly, the
main coolant of halos with a virial temperature of $T_{vir} \approx
10^4 K$ in a strong LW background may continue to remain molecular
hydrogen according to numerical simulations: \citet{oshea08} found
in fact that while in presence of a strong feedback $T_{vir} \approx
10^4 K$ is required for collapse, still the cooling is driven at the
center of the halo by molecular hydrogen, formed thanks to the high
central density (see also \citealt{wise07}). The feedback induced by
nearby Population III sources may also be positive, as, e.g., a soft
X-ray background enhances the $H_2$ production rate
\citep{ric01,macha03}. Therefore the termination of the first epoch of
Population III star formation depends critically on the relative
weight of these two competing process, which in turns is influenced by
the local topology of the IGM, by the spatial distribution of the
sources and by their IMF (which affects the relative efficiency of X
ray to Lyman Werner photon production). In addition to radiative
feedback, Population III star formation is influenced by chemical
feedback. This can be broadly classified as (i) self-enrichment due to
a previous episode of star formation in a progenitor of the halo
considered and (ii) metal pollution due to galactic winds originated
in a nearby halo. Given such a complex scenario it is not surprising
that in the literature there have been many investigations focused on
characterizing the Population III star formation rate and the nature
of the transition from Population III star formed in minihalos to
Population III stars formed in $T_{vir} \approx 10^4$ halos and from
Population III to Population II \citep{mack03,fur05,gre06,smith08}.

The formation of Population III stars is typically investigated by
means of two complementary approaches: (i) analytic models aimed at
deriving an average star formation rate - these usually rely on a dark
matter halo formation rate derived with a \citet{PS} like formalism
combined with recipes to populate the dark halos with Population III
stars \citep{mack03,gre06,wyi06}; (ii) high resolution hydrodynamic -
radiative transfer simulations that follow in detail the collapse and
the early stages of formation of a single Population III star
\citep[e.g. see][]{abel02,oshea07,oshea08,yoshida08}.

In this paper we have two main objectives. First, we focus on the
characterization of the global star formation rate of Population III stars,
thus adopting an analytical approach. We resort to physically
motivated recipes to identify the conditions under which it is
expected that the primordial gas within a dark matter halo can cool
and trigger a gravitational instability which leads to a protostellar
core. These recipes include the effects of a photo-dissociating Lyman
Werner background derived both self-consistently from our model as
well as by adopting a reference reionization history of the
Universe. 

Our second goal is to quantify the probability that a newly formed
dark matter halo with virial temperature $T_{vir} \approx 10^4 K$ has
been previously enriched by one or more episodes of $H_2$ Population
III formation in one of its parent halos. Chemical enrichment of such
halos is in fact crucial not only to assert the relative contribution
of Population III star formation via the atomic and molecular cooling
channels, but also to evaluate the formation rate of quasistars
\citep{beg06} at $z \approx 15$ which have been proposed as
progenitors of the supermassive black holes present after the end of
reionization. Quasistars are in fact able to form only if the gas is
not polluted by metals \citep{omu08}. Our novel approach to
self-enrichment is based on the properties of the Gaussian random
field of the primordial density fluctuations, which allow us to derive
a closed form for the probability that a dark matter halo of mass
$M_1$ at redshift $z_1$ had at redshift $z_2>z_1$ a progenitor of mass
$M_2>M_1$ \citep{ts07a}. We then combine our results on
self-enrichment with the probability of wind pollution derived by
\citet{furl03} to infer the overall likeliness of collapse of pristine
gas in halos with $T_{vir} \approx 10^4 K$.

This paper is organized as follows. In Sec.~\ref{sec:model} we
introduce our model for Population III star formation, including radiative and
self-enrichment feedback; The model is applied in Sec.~\ref{sec:sfr}
to derive the global Population III star formation rate and in
Sec.~\ref{sec:enrich} to obtain the enrichment probability of $T_{vir}
\approx 10^4 K$ dark matter halos. Sec.~\ref{sec:reion} discusses the
implications in terms of contributions to reionization from Population III
stars and Sec.~\ref{sec:conc} concludes. 

\section{Population III star formation model}\label{sec:model}

To derive the star formation rate of Population III stars we combine
the dark matter halo formation rate with an analytical model to
populate halos with stars. In this paper we assume a flat concordance
$\Lambda CDM$ cosmology, with the cosmological parameters given by the
WMAP Yr5 best fitting parameters \citep{kom08}: $\Omega_{\Lambda} =
0.72$, $\Omega_m = 0.28$, $\Omega_b = 0.0462$, $\sigma_8=0.817$, $n_s
= 0.96$, $h=0.7$. We also assume a primordial helium mass fraction
$Y=0.2477$ \citep{helium}.

\subsection{Minimum Minihalo Mass for Population III formation}
 
The minimum dark matter halo mass $M$ capable of cooling by molecular
hydrogen at redshift $z$ is estimated by requiring the cooling time
$\tau_{cool}(M,z)$ to be no larger than the local Hubble time $t_H(z)$
(e.g. see \citealt{cou86}). We write the cooling time as:
\begin{equation}
\label{taucool}
\tau_{cool}(M,z) = \frac{3 k_B T_{vir}(M,z)}{2 \Lambda(T_{vir},n_H) f_{H_2}},
\end{equation}
where $k_B$ is Boltzmann's constant and $T_{vir}(M,z)$ is the virial
temperature of the halo, $\Lambda$ is the cooling function per $H_2$
molecule, which depends on the temperature and on the hydrogen number
density $n_H$ and $f_{H_2}$ is the molecular to atomic hydrogen
fraction. We write $T_{vir}$ following \citet{teg97} as:
\begin{equation}
\label{Tvirial}
T_{vir}(M,z) \simeq 2554 K \left(\frac{M}{10^6 \mathrm{M_{\sun}}}\right)^{2/3}
\left(\frac{1+z}{31}\right).
\end{equation}
For the molecular hydrogen cooling function $\Lambda$, we use the form
derived by \citet{galli98}, which we approximate between the
temperatures of 120 K and 6400K (the range we are interested in for
cooling in minihalos) with:
\begin{equation}
\label{lambda}
\Lambda(T, n_H) \simeq 10^{-31.6} \times \left(\frac{T}{100 K}\right)^{3.4} \times \left(\frac{n_H}{10^{-4} \mathrm{cm^{-3}}}\right) \mathrm{erg s^{-1}}.
\end{equation}
We estimate the hydrogen number density in a halo from its virial density
(e.g. Eq.~22 of \citealt{teg97}) to find:
\begin{equation}
\label{hydrogendensity}
n_H \simeq 1.01 \left(\frac{1+z}{31}\right)^3 \mathrm{cm^{-3}}.
\end{equation}
Replacing Eqs. \ref{Tvirial}, \ref{lambda}, \ref{hydrogendensity} in
Eq. \ref{taucool}, we find:
\begin{equation}
\label{eq:taucoolofMz}
\tau_{cool} \simeq 3.46 \times 10^{10} {\rm s} \left(\frac{M}{10^6 M_\odot}\right)^{-1.6} \left(\frac{1+z}{31}\right)^{-5.4} f_{H_2}^{-1}.
\end{equation}
We can then obtain the molecular hydrogen fraction required for cooling
by equating the cooling time given by Eq. \ref{eq:taucoolofMz} to the
local Hubble time $t_H$, approximated at $z \gg 1$ as:
\begin{equation}\label{eq:th}
t_H(z) \simeq \frac{1.191 \times 10^{15} {\rm s}}{h \sqrt{\Omega_m}} \left(\frac{1+z}{31}\right)^{-3/2} \simeq 3.216 \times 10^{15} {\rm s} \left(\frac{1+z}{31}\right)^{-3/2}.
\end{equation}
This gives us:
\begin{equation}
\label{minimumh2frac}
f_{H_2} \simeq 1.09 \times 10^{-5} \left(\frac{M}{10^6 M_\odot}\right)^{-1.6}
\left(\frac{1+z}{31}\right)^{-3.9}.
\end{equation}
\citet{teg97} determine the (maximum) molecular hydrogen fraction capable of forming in a halo as: 
\begin{equation}\label{eq:fh2_max}
f_{H2,max} \simeq 3.5 \times 10^{-4} \left(\frac{T}{1000K}\right)^{1.52}.
\end{equation}
Equating the required molecular hydrogen fraction for cooling within a
Hubble time given by Eq. \ref{minimumh2frac} with the maximum that can
form (assuming $T=T_{vir}$) we find the minimum mass 
for a minihalo in order to cool within a Hubble time ($M_{t_H-cool}$), namely:
\begin{equation}
\label{minimummass}
M_{t_H-cool} \simeq 1.54 \times 10^5 M_\odot
\left(\frac{1+z}{31}\right)^{-2.074}.
\end{equation}

\subsection{Minimum $H_2$ cooling mass in presence of radiative feedback}

In order to compute the effect of a radiative flux in the Lyman-Werner
band on the formation rate of molecular hydrogen and on the cooling of
a minihalo, we resort to an approach based on \citet{macha03}. We
obtain the minimum halo mass capable of cooling via $H_2$ in the
presence of a LW background by equating the timescale for
photo-dissociation of molecular hydrogen ($\tau_{diss}$) to its
formation timescale ($\tau_{form}$).

In presence of a LW flux $F_{LW} = 4 \pi J_{21} 10^{-21} \mathrm{erg
s^{-1} cm^{-2} Hz^{-1}}$ (whose calculation is presented below in
Sec.~\ref{sec:LW_flux}), the dissociation timescale can be written as
\citep{macha03}:
\begin{equation}
\label{taudiss}
\tau_{diss} \simeq \frac{7.16 \times 10^{11} s}{J_{21}},
\end{equation}
The $H_2$ formation time scale is given by:
\begin{equation}
\label{tauform}
\tau_{form} = \frac{n_{H_2}}{k_7 n_H n_e} \simeq \frac{f_{H_2}}{k_7 n_e},
\end{equation}
where $k_7 \approx 1.8 \times 10^{-18} T^{0.88}$ cm$^3$ s$^{-1}$ is
the H$^-$ formation rate which dominates the formation of molecular
hydrogen and $n_e$ is the electron density. $n_e$ is obtained assuming a residual ionizing fraction $2 \times 10^{-4}$ \citep{peebles}. Imposing the equilibrium
molecular hydrogen function to be the minimum needed for collapse as
given by Eq. \ref{minimumh2frac} we finally find:
\begin{equation}\label{eq:m_h2_cool}
M_{H_2-cool} \simeq 6.44 \times 10^6 M_\odot J_{21}^{0.457}
\left(\frac{1+z}{31}\right)^{-3.557}.
\end{equation}
We note that Eq.~\ref{eq:m_h2_cool} is in good agreement with the
results by \citet{macha03} at $z \simeq 30$. Our derivation does
however include an explicit redshift
dependence. Eq.~\ref{eq:m_h2_cool} also compares well with the results
of numerical simulations by \citet{oshea08}, which include LW
background of varying intensities. The redshift dependence which we
find increases the minimum mass required for cooling with respect to
the formula by \citet{macha03} at $z < 30$ and this gives us a better
agreement with the numerical results, obtained for $z \lesssim 25$
(see Fig. 3 in \citealt{oshea08}).

In conclusion, for a dark matter halo to be able to cool via $H_2$,
its mass must be above both the limits set by Eq.~\ref{minimummass}
and Eq.~\ref{eq:m_h2_cool}, that is
\be \label{eq:m_minihalo}
M_{min} = max(M_{t_H-cool};M_{H_2-cool}).
\ee

\subsection{Cooling in halos with $T_{vir}\gtrsim 10^4 K$}\label{sec:10^4Khalos}

Pristine halos with a virial temperature above $T_{vir} \simeq10^4 K$
can cool irrespective of the LW background intensity
\citep{oshea08}. In fact, not only atomic hydrogen cooling becomes in
principle available, but also cosmological simulations by
\citet{oshea08} have shown that a small fraction of $H_2$ can still be
produced at the center of the halo thanks to the high density and
self-shielding of the surrounding gas. Once the gas temperature starts
to decrease, further cooling and collapse will proceed progressively
faster via molecular hydrogen as the halo temperature is initially
high enough to enhance the abundance of $H^{-}$, a precursor for $H_2$
production \citep{lepp84}. In our model we thus consider that all
halos above the $M_{T=10^4 K}$ limit will cool efficiently:
\be
M_{T=10^4 K} =  7.75 \cdot 10^6  \mathrm{M_{\sun}} \left ( \frac{1+z}{31} \right )^{-3/2}.
\ee

\subsection{Forming Population III stars}

There is of course a delay between the virialization of a dark matter
halo potentially able to cool via $H_2$ --- that is of mass $M >
M_{min}$ (Eq.~\ref{eq:m_minihalo}) --- and the actual formation of a
Population III star. We estimate this delay considering two contributions: (i)
the actual time needed to cool down to a few hundreds degrees Kelvin
and (ii) the free fall time for the gravitational collapse once
cooling has triggered the Jeans instability.

The $H_2$ cooling time can be obtained by
Eqs.~\ref{eq:taucoolofMz} and \ref{eq:fh2_max}:
\be
\tau_{cool} = 2.38 \times 10^{13} \mathrm{s} \left ( \frac{M}{10^6 \mathrm{M_{\sun}}}
\right )^{-2.627} \left ( \frac{1+z}{31} \right )^{-6.94}, 
\ee
while the free fall time can be obtained from the Jeans instability
timescale, taking into account that during the cooling phase the
density of the gas has increased by about a factor $7$:
\be
t_{ff} =  2.77 \times 10^{14} \mathrm{s} \left ( \frac{1+z}{31} \right )^{-3/2}.  
\ee

Therefore a Population III star originated in dark halo virialized at redshift
$z_{vir}$ will be formed at redshift $z_{form}<z_{vir}$ such that:
\be
t_{cool}+t_{ff} = 3.21 \times 10^{15} {\rm s} \left [\left( \frac{1+z_{form}}{31} \right )^{-3/2} - \left ( \frac{1+z_{vir}}{31} \right )^{-3/2} \right ],
\ee
where the right side of the equation simply derives from the age of
the Universe at $z \gg 1$ (Eq.~\ref{eq:th}).

\subsection{Metal enrichment probability}\label{sec:metal_enrich}

In order to account for previous episodes of Population III star formation
in a progenitor of a halo of mass $M$ at redshift $z_1$, we resort to
the method presented in \citet{ts07a}, based on the linear growth of
density perturbation in the context of spherical collapse.  We start
by assuming that a newly virialized halo has an average linear
overdensity $\delta(z)=1.69$ as estimated by a top-hat filter on a
scale $R = (3M / 4 \pi \langle \rho \rangle)^{1/3}$. Then for a
progenitor mass $M_{prog}<M$ we compute the extra variance in the
density power spectrum $\sigma^2_{add} =
\sigma^2(M_{prog})-\sigma^2(M)$ and then the refinement factor
$N_{ref} = M/M_{prog}$. With these ingredients we can write the
probability distribution for the maximum of $N_{ref}$ Gaussian random
numbers with variance $\sigma^2_{add}$ as the derivative of the
$N_{ref}$ power of the Partition function for a normal distribution
with zero mean and variance $\sigma^2_{add}$. In the context of
spherical collapse this translates to a probability distribution for
the formation redshift of the first progenitor of mass $M_{prog}$ of a
halo $M$ virialized at $z_1$.

For every progenitor mass $M_{prog}$ we then compute the delay time
($t_{ff}+t_{cool}$) needed to form a Population III star in the parent
halo and from this we derive the minimum redshift ($z_{min\_seed} $)
at which such a halo must form in order to pre-seed the descendant
halo. Of course, for some values of $M_{prog}$ the delay time might be
longer than the Hubble time, this simply means that no pre-seeding is
possible from parent halos of mass $M \leq M_{prog}$. We then
integrate the probability distribution for the formation time of the
parent of mass $M_{prog}$ for $z>z_{min\_seed}$ to obtain the
preseeding probability from a progenitor at this mass scale. The
overall probability of preseeding is the maximum preseeding
probability computed over all the possible progenitor masses.

\subsection{From dark matter to stars}

The dark matter halo formation rate is derived in our reference model
using the \citet{she99} mass function. The \citet{she99} mass function
is in better agreement with N-body simulations than the \citet{PS}
mass function at $z \lesssim 30$ \citep{heit06,reed07}. Note that
differences of $\approx 20\%$ have been observed between the
measurements from cosmological simulations and the \citet{she99} mass
function and that the \citet{warren06} mass function appears a better
match to the numerical results \citep{lukic07}. However the
\citet{she99} and the \citet{warren06} formulae give very similar
results in the range of halo masses of interest for Population III
star formation ($M \lesssim 10^{8} \mathrm{M_{\sun}}$ --- see Fig.~3
in \citealt{lukic07}) thus we keep the \citet{she99} model as our
reference. We then compute the formation rate of $H_2$ Population III
stars by integrating between $M_{min}(z)$ and $M_{T=10^4 K}(z)$ the
number of halos per unit mass per unit redshift $dN(M,z)/dM dz$,
convolved with the probability that such halos are pristine (see
Sec.~\ref{sec:metal_enrich}). 

The characteristic mass of Population III stars and the form of their
initial mass function are highly uncertain, even though they are
likely very massive --- of the order of $O(100 \mathrm{M_{\sun}})$ ---
\citep[e.g. see][]{bromm04}. This expectation is based on theoretical
models and numerical simulations
\citep{abel02,omukai03,yoshida06,gao07}, but there is some tension
with the abundance patterns observed in the most metal poor Milky Way
stars, which are better explained under the assumption that their
progenitor Population III stars were only moderately massive --- $8
\mathrm{M_{\sun}} \lesssim M \lesssim 42 \mathrm{M_{\sun}}$ ---
\citep{tum06}. There is however no guarantee that the progenitors of
the extremely metal poor stars considered by \citet{tum06} are
Population III stars formed before the reionization of the Universe:
if their progenitors formed instead in presence of a strong UV
background within a reionized bubble, then their expected mass is
about $\approx 40 \mathrm{M_{\sun}}$ fully consistent with the
inference from the observations \citep{yoshida07}. Within this complex
scenario we choose to adopt conventionally one Population III per
minihalo \citep{oshea07} and we consider a \citet{sal} mass function
in the range $[50:300] \mathrm{M_{\sun}}$ (average mass $100
\mathrm{M_{\sun}}$), as suggested by the theoretical investigations. A
modification in the initial mass function used in our model primarily
affects the enrichment history of the IGM and thus the transition
toward Population II star formation. If Population III stars are less
massive than we assume, then the efficiency of metal pollution may be
reduced as core collapse supernovae explosions are not as energetic as
pair instability ones \citep{heger03}. A mass function more biased
toward very massive stars with $M>270 \mathrm{M_{\sun}}$ would also
reduce the efficiency of metal pollution, because these stars directly
collapse into black holes without an explosive phase \citep{heger03}.

The formation rate for Population III stars in halos with $T_{}>10^4
\mathrm{K}$ is similarly computed from the dark matter halo mass
function for $M>M_{T=10^4 K}$, again after convolution with the
probability that the gas forming new halos has not been contaminated
by metals. As no sign of fragmentation has been found during the
collapse of metal free halos with masses up to $2 \times 10^{7}
\mathrm{M_{\sun}}$ \citep{oshea08}, we adopt the same initial mass
function as for $H_2$ cooled Population IIIs (one star per halo,
Salpeter in $[50:300] \mathrm{M_{\sun}}$). The efficiency of star
formation is however uncertain and thus we explore different models
where multiple Population III stars in a single halo are allowed,
adopting a reference star formation efficiency (star to gas mass
ratio) of $\epsilon_{PopIII} = 0.005$ and $\epsilon_{PopIII} = 0.05$.

Finally the star formation rate in enriched gas (Population II stars)
is computed by convolving the dark matter halo mass function with the
preseeding probability (the complementary of the pristine probability)
and assuming a star formation efficiency of $\epsilon_{PopII} = 0.005$
or $\epsilon_{PopII} = 0.05$ (to explore the uncertainties in this
parameter), a Salpeter mass function from in the range $[1:100]
\mathrm{M_{\sun}}$ (average mass $\approx 3 \mathrm{M_{\sun}} $) and an
average metallicity $10^{-4} Z_{\sun}$. Our choice of the IMF for
metal enriched gas reflects the expectation that the typical mass of
stars was higher at higher redshift \citep{tum07}. In our model, the
IMF of metal enriched star formation impacts the radiative LW
feedback, but only in a minor way due to its self-regulating nature
(see Sec.~\ref{sec:sfr}).

\subsection{Flux in the LW band}\label{sec:LW_flux}

We compute the LW flux that enters in Eqs.~\ref{taudiss} by means of
two different approaches: (i) self-consistently from our model, based
on the star formation rate and (ii) adding a reference number of LW
photons to those derived self-consistently in order to take into
account Population II formation not included in our model. The first
method is most suitable at $z \gtrsim 20$, when Population III are
most likely the dominant sources of radiation. At lower redshift
protogalaxies become more and more common, the Universe starts
becoming reionized and our simple model for star formation does not
capture all the Population II formation that is available, therefore
using a reference LW photons production provides a good check on the
validity of our assumed LW flux.

For the self-consistent LW flux calculation, we obtain that Population
III stars following our assumed IMF emit a LW flux that is $7.5
\%$ of the ionizing flux \citep{schae03}. Metal enriched stars have
instead a higher ration of LW to ionizing photons, because these stars
have a lower effective temperature, but their LW photon yield per unit
mass is also lower. We assume the following photon yields over the
star lifetime (based on \citealt{schae03}):
\begin{enumerate}
\item Population III: $8 \times 10^{60} M_{\sun}^{-1}$;
\item Population II:  $8 \times 10^{59} M_{\sun}^{-1}$;
\end{enumerate}
The comoving LW photon density $n_{LW}$ is then associated to a flux:
\begin{equation}
J_{21} = 1.6 \times 10^{-65} \left ( \frac{n_{LW}}{\mathrm 1 Mpc^{-3}} \right ) \left ( \frac{1+z}{31} \right)^3 \mathrm{erg s^{-1} cm^{-2} Hz^{-1} sr^{-1}}.
\end{equation}
In order to compute $n_{LW}(z)$ we only consider star-formation that
has happened within a redshift interval such that the photons have not
been redshifted out of the LW band on average. This means that the
upper limit, expressed in term of the redshift is $z_{up} = 12.39 /
11.18 (1+z)-1$. We also take into account the screen provided by
primordial $H_2$ present in the IGM outside virialized halos, which
can absorbs LW photons. Following \citet{ts07a} a flux of $J_{21} =
1.58 \times 10^{-3}$ is needed to photo-dissociate a molecular
hydrogen density of $10^{-6}$ times the neutral hydrogen density. Of
course, a (very) small residual fraction of $H_2$ will still be
present but the re-formation of $H_2$ outside virialized structures is
strongly suppressed not only by the radiative feedback, but also by
the decreased average density of the universe compared to that at the
time of primordial $H_2$ formation ($z\approx 200$ --- see
\citealt{peebles}).

For the LW flux based on a fixed reionization history of the Universe,
we assume that the number of LW photons is that produced
self-consistently by Population III plus a contribution from other
sources which reaches $n_{LW} = 7 \times 10^{66}$ at $z=10$
(corresponding to $J_{21} \approx 4.9$). As a model for the redshift
dependence of $n_{LW}$ we take inspiration from the rapid growth of
the fraction of mass in collapsed dark matter halos that can host
stars and we write:
\begin{equation}\label{eq:ext_LW}
n_{LW}(z) =  7 \times 10^{66} \times 10^{3.3-3.3 \times (1+z)/11}.
\end{equation}
 The flux in the LW band approximately matches the reionizing flux
given for a stellar population with a Salpeter IMF in the range
$[1:100] \mathrm{M_{\sun}}$ and metallicity $Z\approx 10^{-3}
Z_{\sun}$ \citep{schae03}.  Thus Eq.~\ref{eq:ext_LW} implies that
about one ionizing photon per hydrogen atom is emitted at redshift
$10$, a budget still well short of what is required for reionization
after considering the clumpiness of the IGM and recombination.

\section{Population III star formation rates}\label{sec:sfr}

\begin{figure} 
  \plottwo{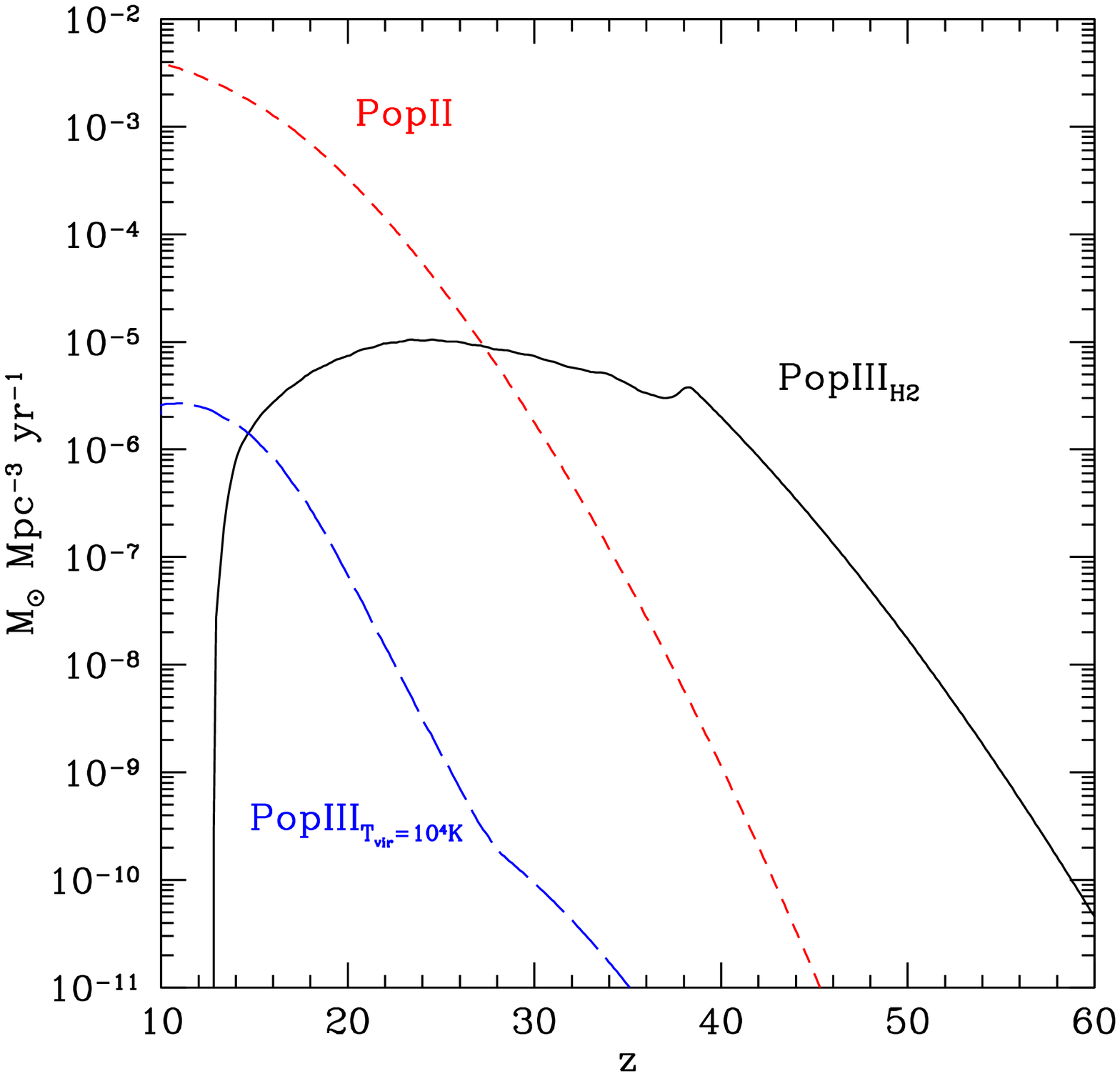}{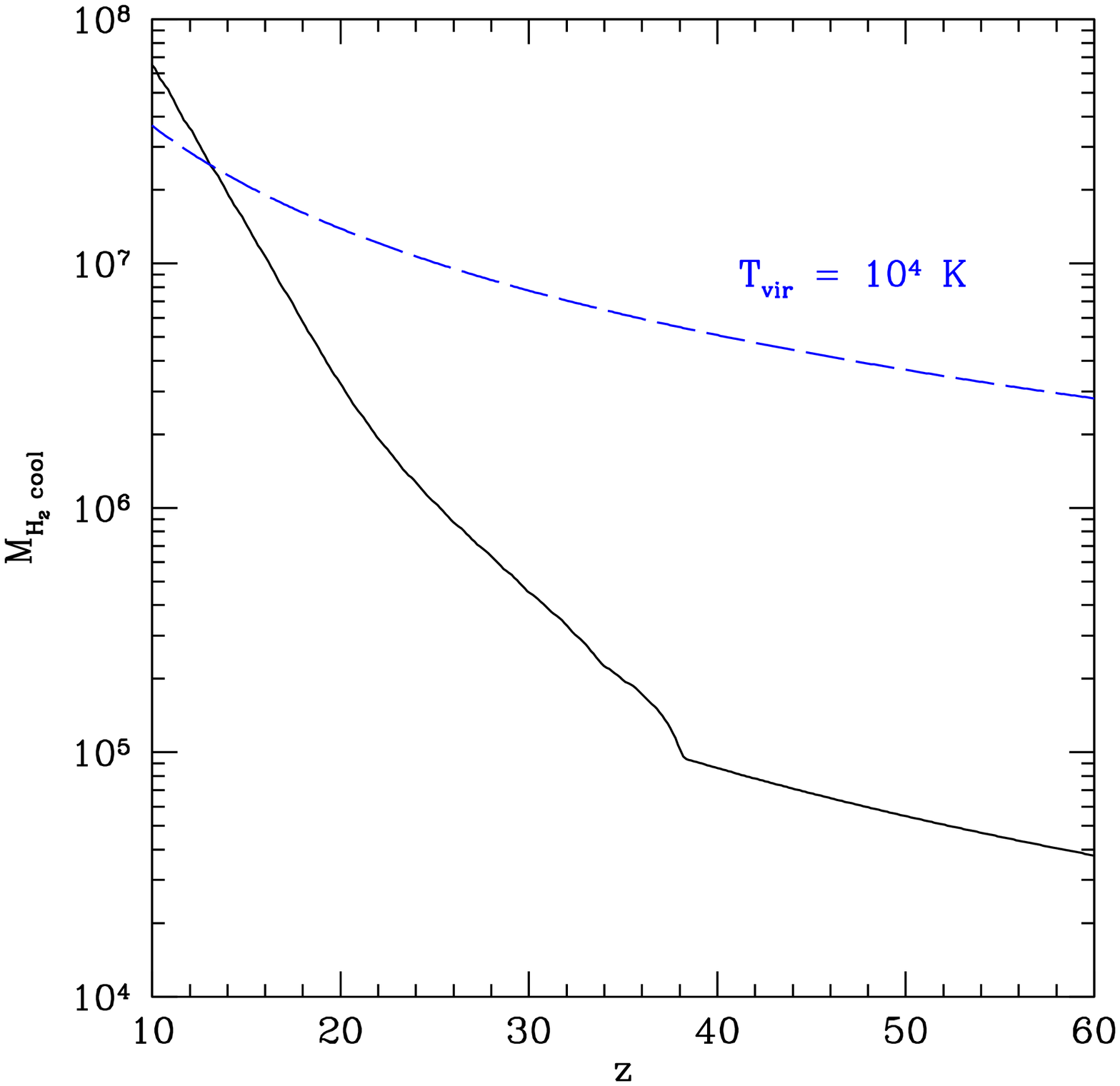}
  \plottwo{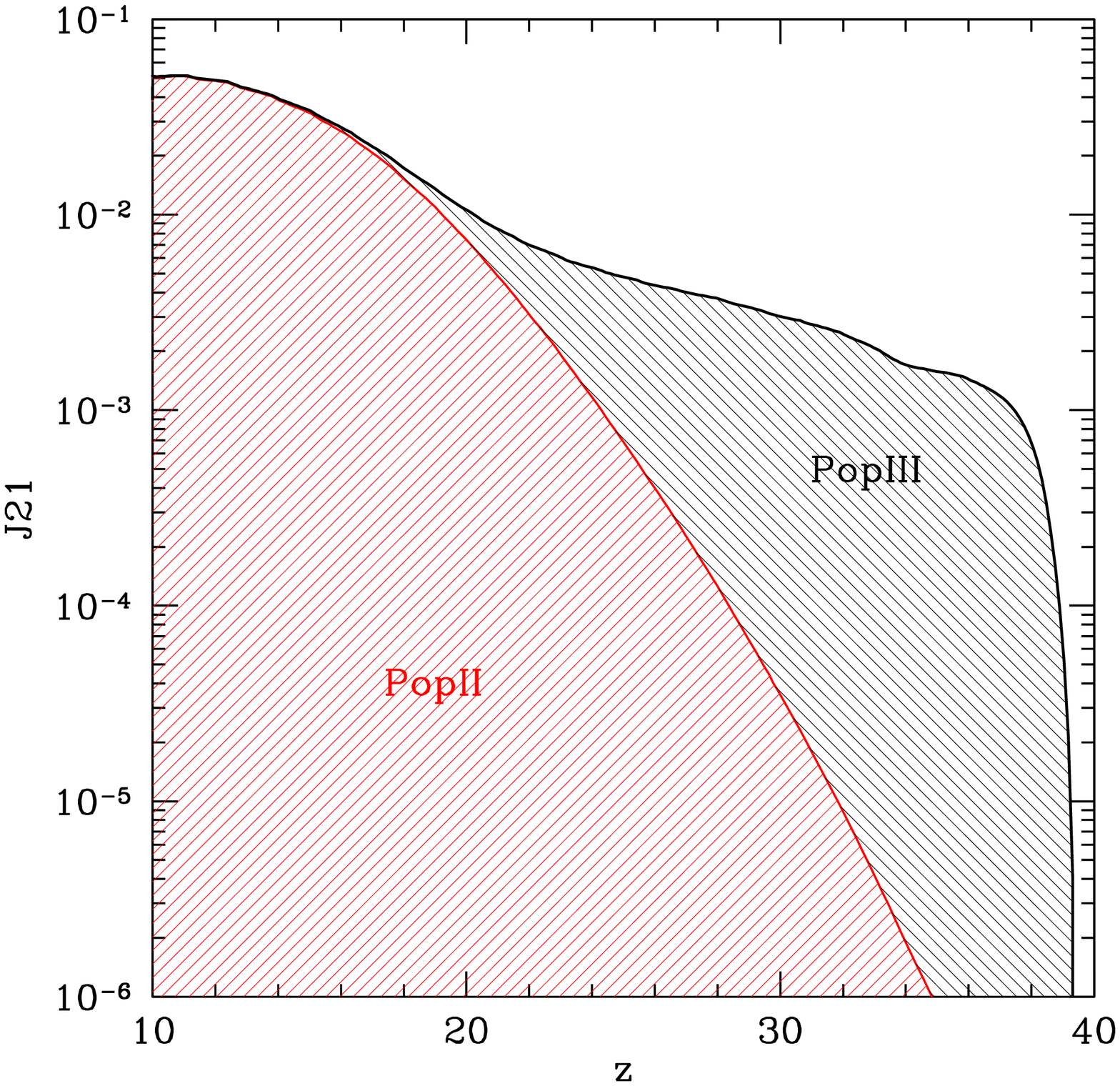}{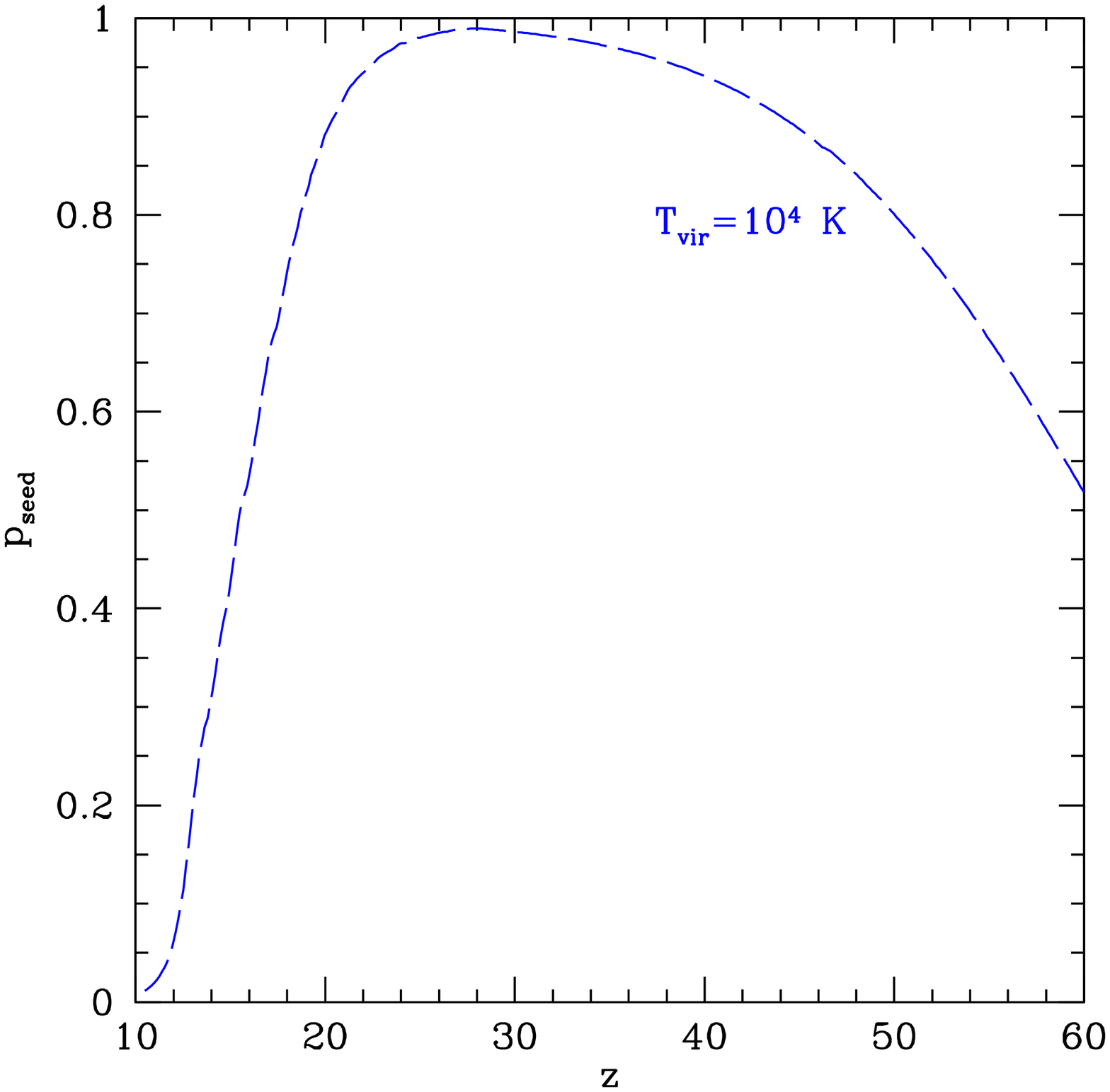}
\caption{Star Formation during the reionization epoch predicted by our
standard model for Population III stars. Upper left panel: star
formation rate versus redshift for Population III stars in minihalos
(solid black line), Population III stars in more massive halos, with
$T_{vir}\geq 10^4 \mathrm{K}$ (long dashed blue line) and for
Population II stars formed out of metal enriched gas (short dashed
red line). Upper right panel: minimum dark matter halo mass required
to form a Population III star via $H_2$ cooling (solid black line) and
in halos with $T_{vir} \geq 10^4 K$ (long dashed blue line) in function of
redshift. Lower left panel: $J_{21}$ flux in function of redshift with
contribution from Population III stars (black shaded area) and from
Population II stars (red shaded area). For $z\gtrsim 22$
Population III stars are the main source of radiative feedback. Lower
right panel: Probability of metal enrichment via progenitor pollution
for a halo with $T_{vir}=10^4 \mathrm{K}$. The results have been
obtained using a WMAP5 cosmology, a \citet{she99} halo mass function
and our model includes cooling and J21 feedback. The star formation
rate is one star per halo for Population III stars and $5 \times
10^{-3}$ for Population II stars.}\label{fig:popIII_sfr_std}
\end{figure}

From our fiducial model, which has one first star per halo (see
Table~\ref{tab:params} for a summary of the main parameters), it is
immediate to note that Population III stars initiate the chemical
enrichment of the Universe well before redshift $z=50$ (see
Fig.~\ref{fig:popIII_sfr_std}). This result derives from the
relatively small mass required at very high redshift to be able to
cool via molecular hydrogen. Such mass is in fact as low as $M \approx
4 \times 10^{4} M_{\sun}$ at $z=60$. At lower redshift the minimum
mass for cooling progressively increases but halos capable of cooling
become more and more common, so the Population III star formation rate
steadily increases. Eventually --- around redshift $z\approx 35$ ---
radiative feedback in the LW band starts to significantly increase the
mass required for cooling and the star formation rate of this class of
Population III stars levels off at $\approx 10^{-5} \mathrm{M_{\sun}
Mpc^{-3} yr^{-1}}$, that is about one star in a comoving Mpc$^{3}$
formed per unit redshift. In our reference model the self-shielding
mass becomes larger than that of a $T_{vir}=10^4 K$ halo at $z \approx
13$. Following Sec.~\ref{sec:10^4Khalos} we assume that such halos can
cool independently of the LW background intensity.

Forming Population III stars in these more massive halos is possible
only if there were no previous episode of star formation within their
progenitors. Thus, as long as cooling can be efficiently achieved via
$H_2$, halos with $T_{vir}>10^4 \mathrm{K}$ are most likely chemically
enriched (see bottom right panel of Fig.~\ref{fig:popIII_sfr_std}) and
Population III stars in these halos are very rare compared to their
counterparts in minihalos (see the upper left panel of the same
figure). Their star formation rate becomes higher than that in
minihalos only at $z \lesssim 14$, when the LW feedback strongly
suppresses $H_2$ cooling and thus it is more likely that a halo made
entirely of pristine gas is able to grow via mergers to reach
$T_{vir}>10^4$ before having the possibility to cool.

The majority of halos with $T_{vir}>10^4 \mathrm{K}$ at $z \gtrsim 16$
have been instead chemically enriched and Population II star formation
grows rapidly in time. By $z \approx 26$ it becomes the dominant
factor in the global star formation rate, despite our conservative
assumption that only $0.5 \%$ of the gas is converted into stars (but
note that this is still larger than the efficiency in Population III
stars at $z\lesssim 30$ if only one per halo is formed). This early
rise of Population II is a novel result which derives from our
detailed treatment of self-enrichment. When metal transport is instead
modeled via winds, then transition toward Population II stars is
predicted significantly later, at $10 \lesssim z \lesssim 20$
\citep{furl03,fur05}.

\begin{figure} 
  \plottwo{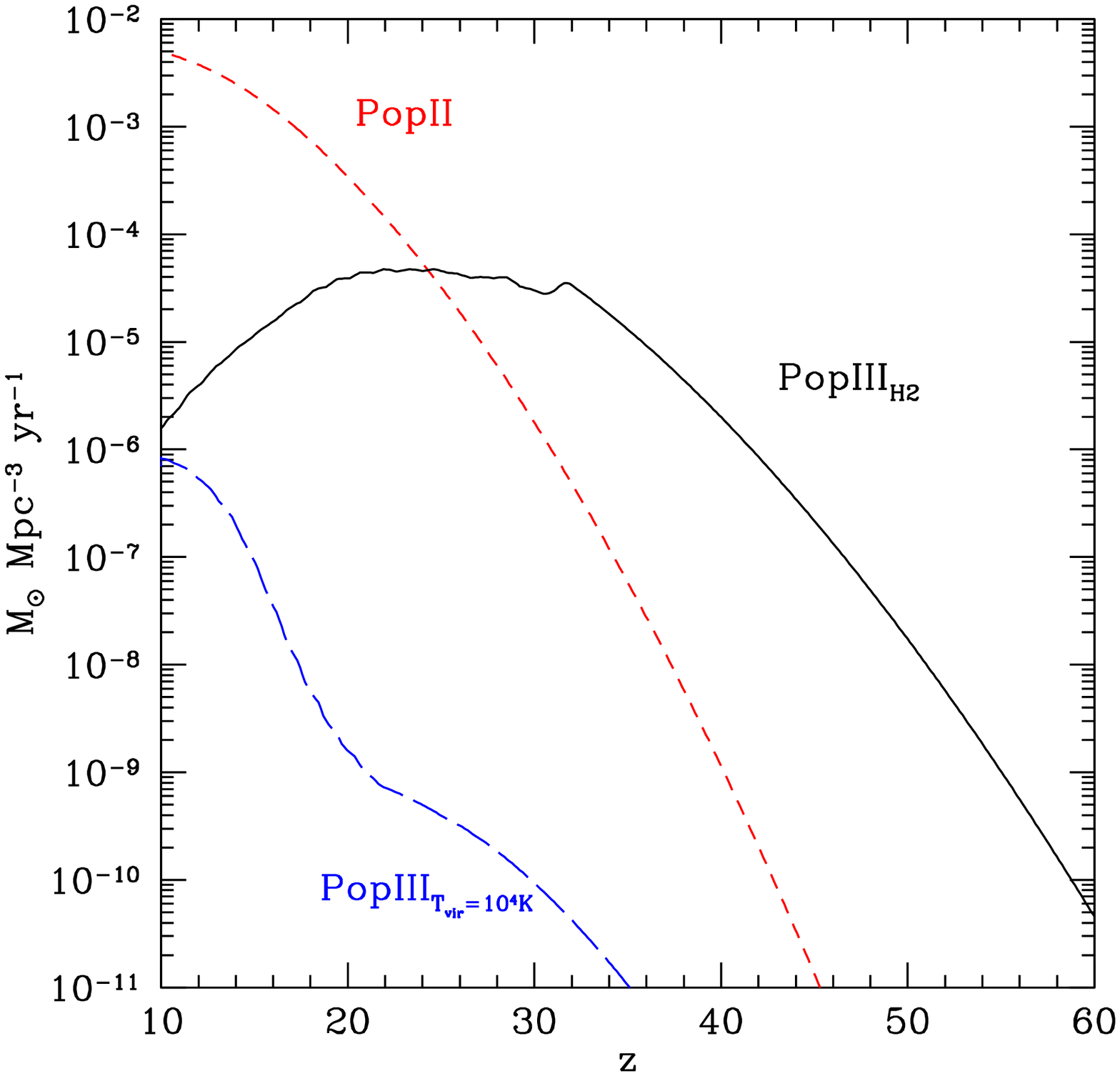}{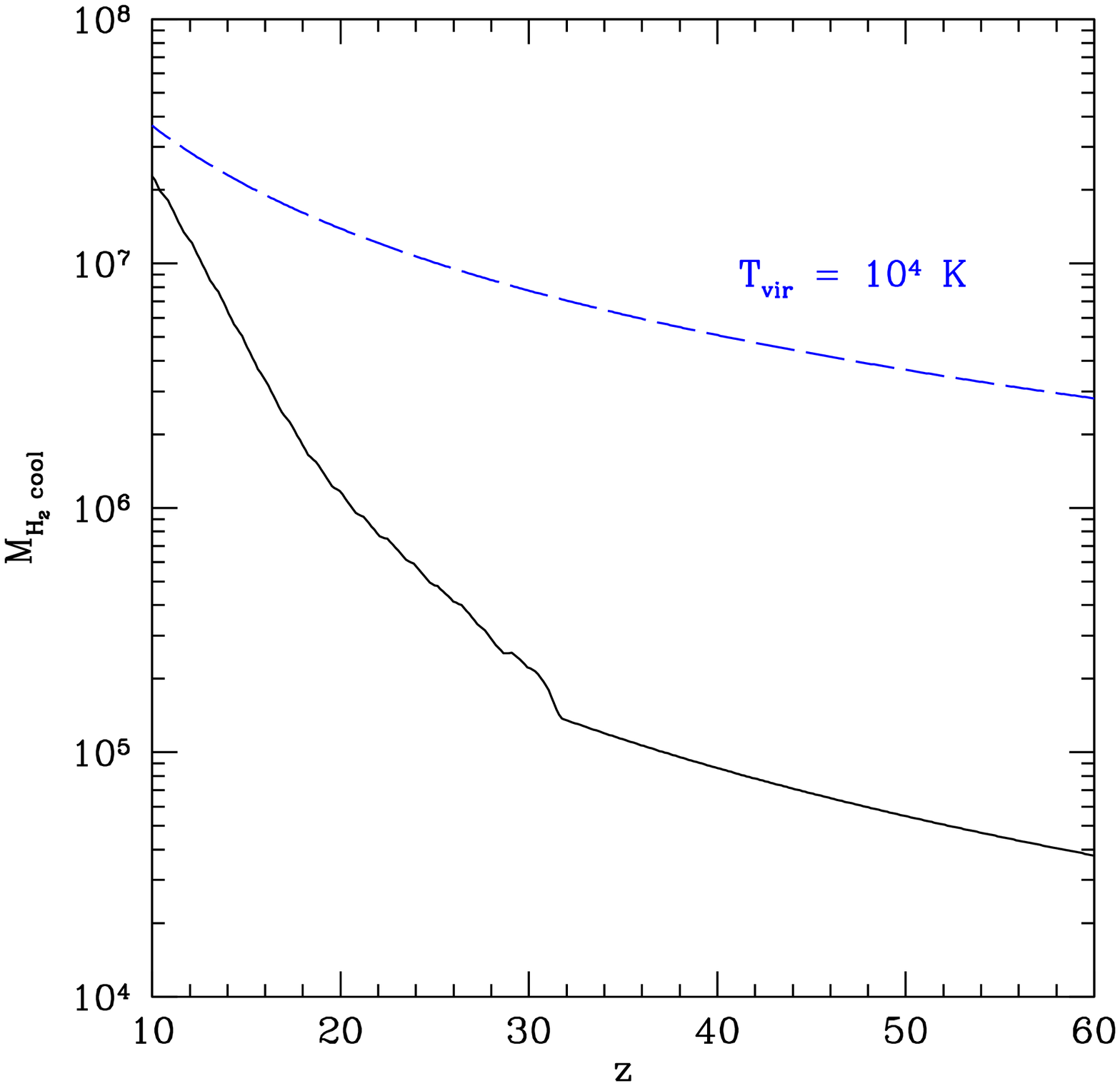}
  \plottwo{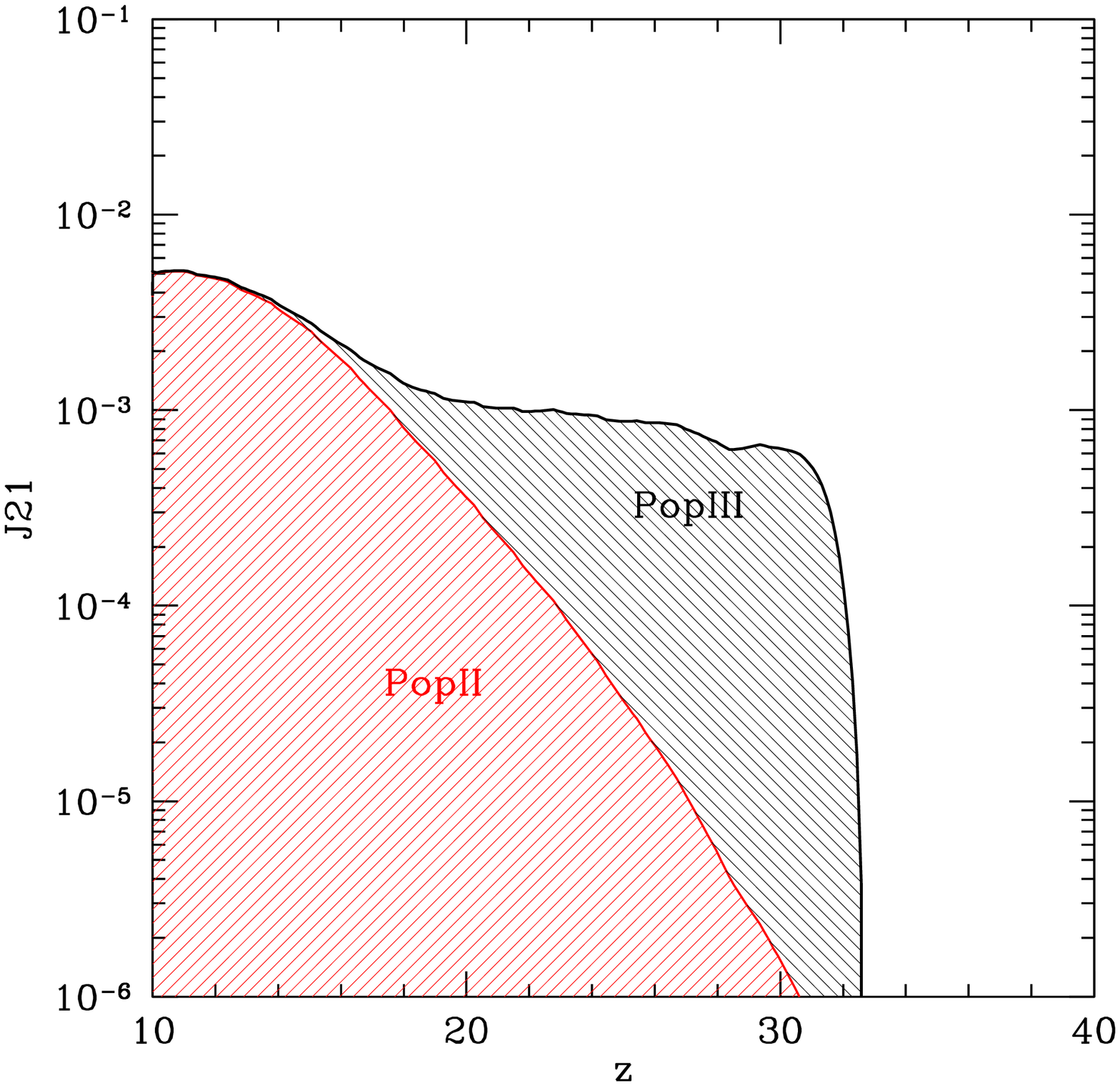}{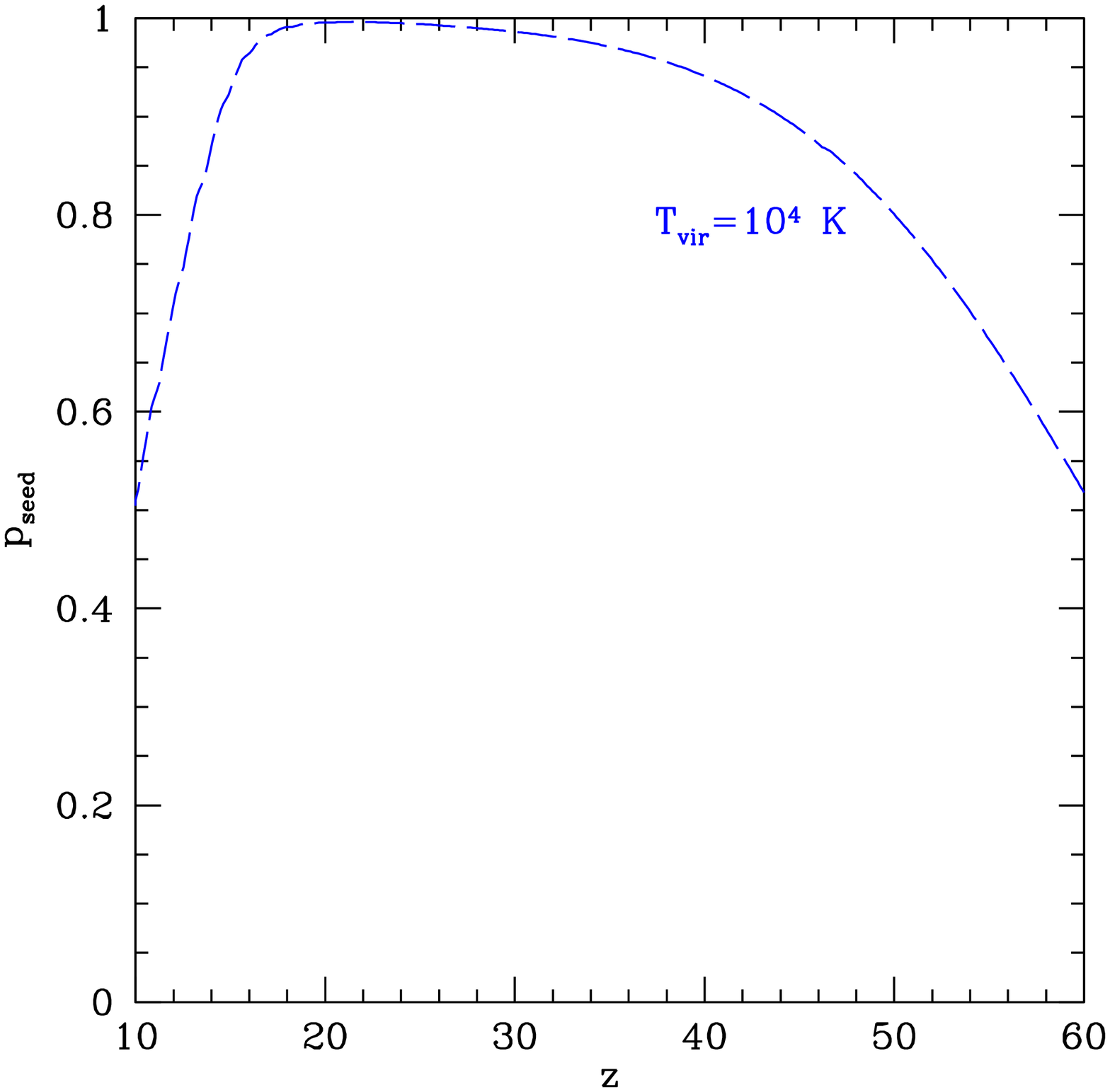}
\caption{Like in Fig.~\ref{fig:popIII_sfr_std} but for our model with
a reduced escape fraction, which implies a less efficient LW
feedback. In this case the formation rate of $H_2$ Population III
stars is increased, while Population III stars in halos with $T_{vir}
\geq 10^4 K$ are partially quenched by chemical enrichment compared to
fig.~\ref{fig:popIII_sfr_std}.}\label{fig:low_escape}
\end{figure}

\begin{figure} 
  \plottwo{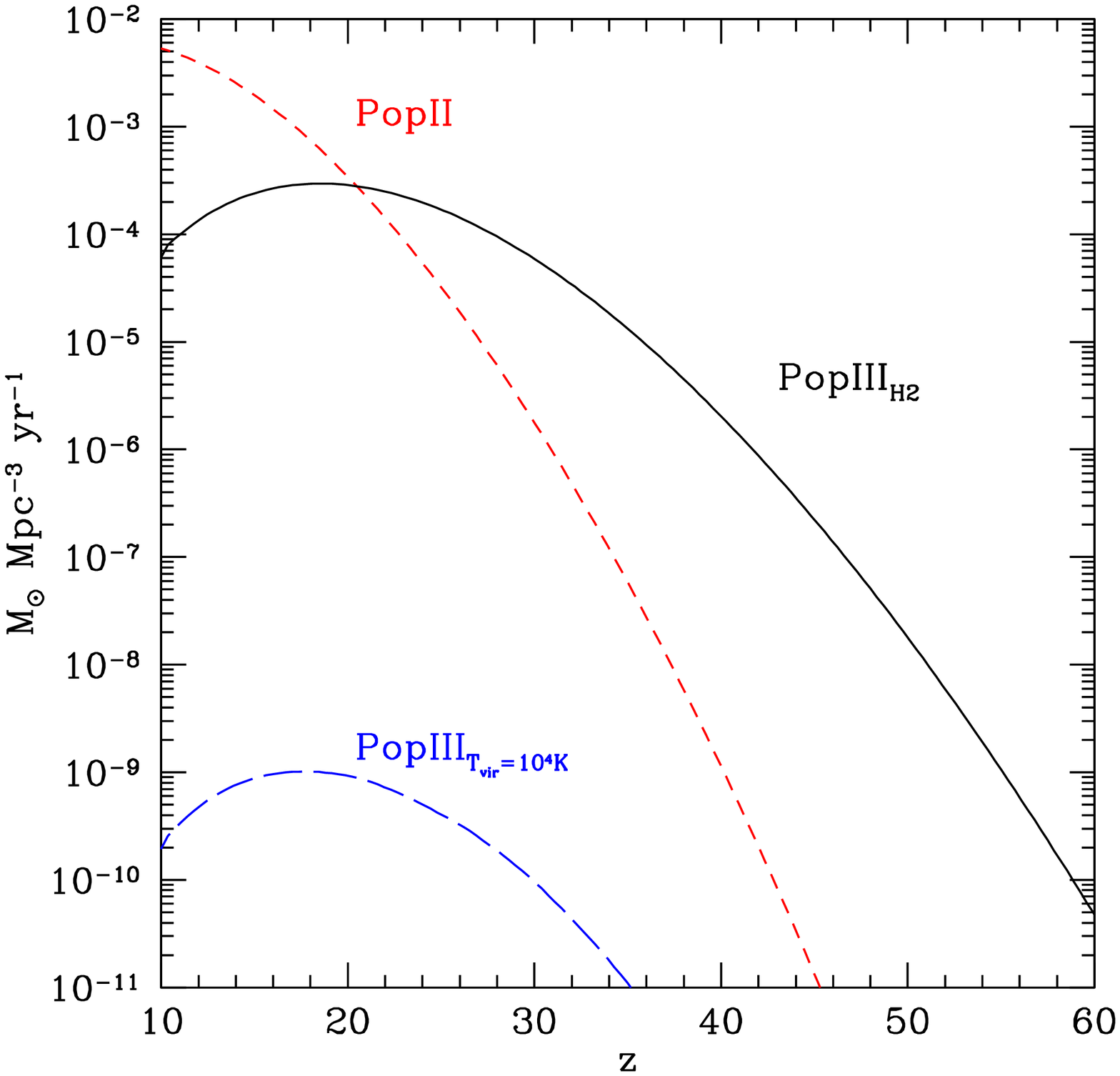}{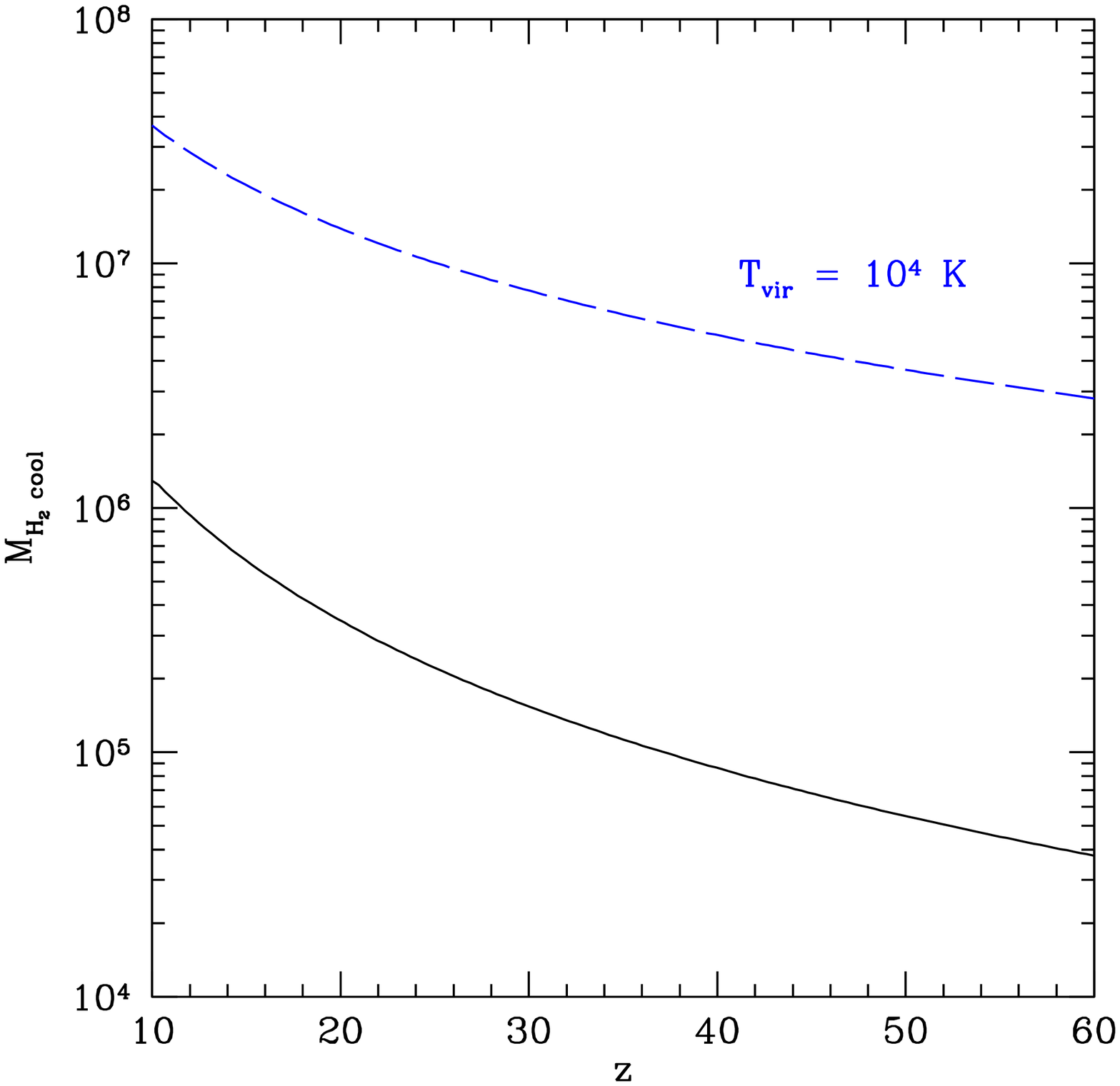}
  \plottwo{fig3_c.eps}{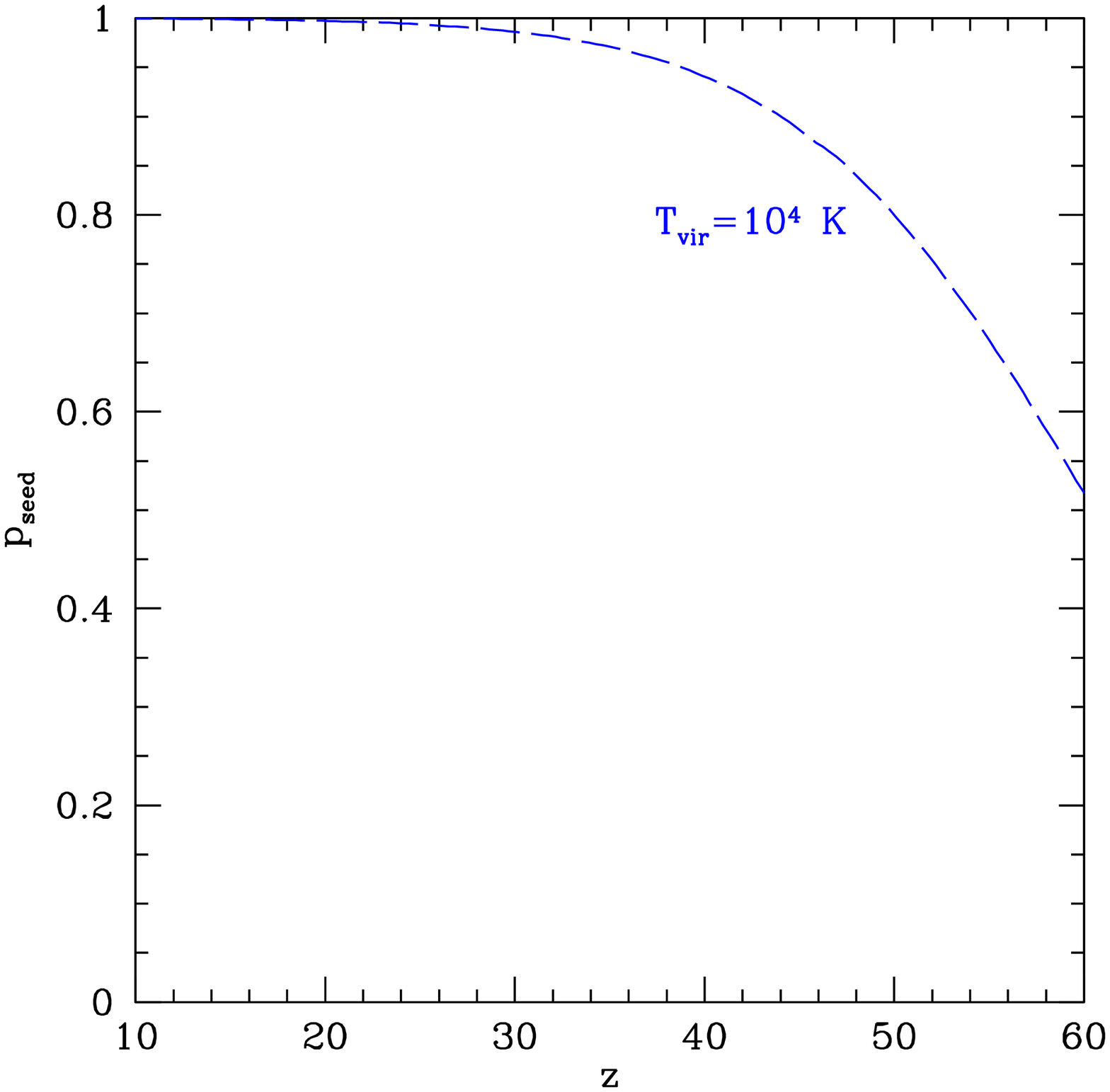}
\caption{Like in Fig.~\ref{fig:popIII_sfr_std} but for our model with
no LW feedback. Note that in this model $J_{21}$ does not influence
star formation, hence it is not shown. In this model quenching of
Population III stars in the more massive $T_{vir} \geq 10^4 K$ halos
is further enhanced compared to Fig.~\ref{fig:low_escape}.}\label{fig:no_feedback}
\end{figure}

\begin{figure} 
  \plottwo{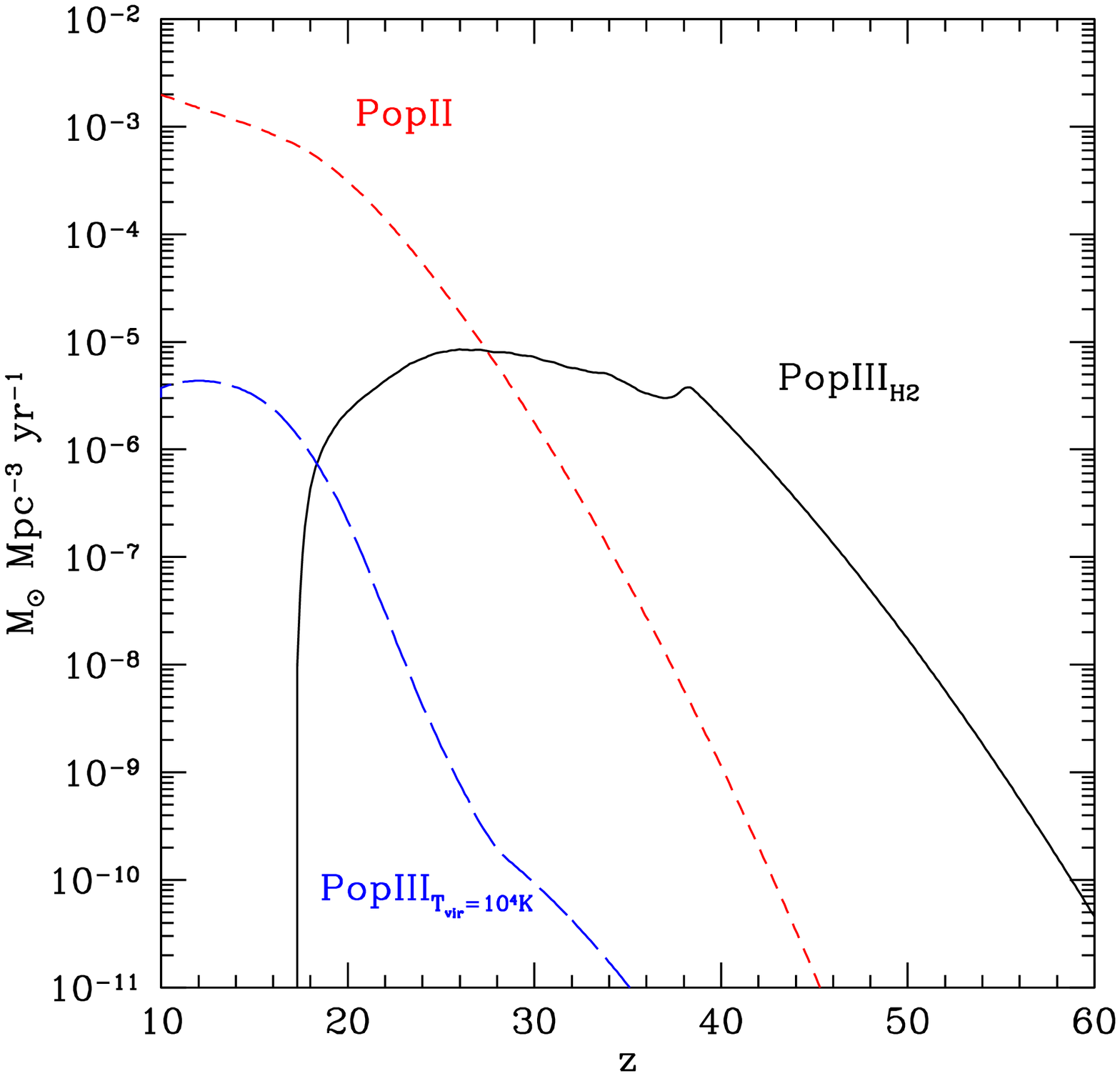}{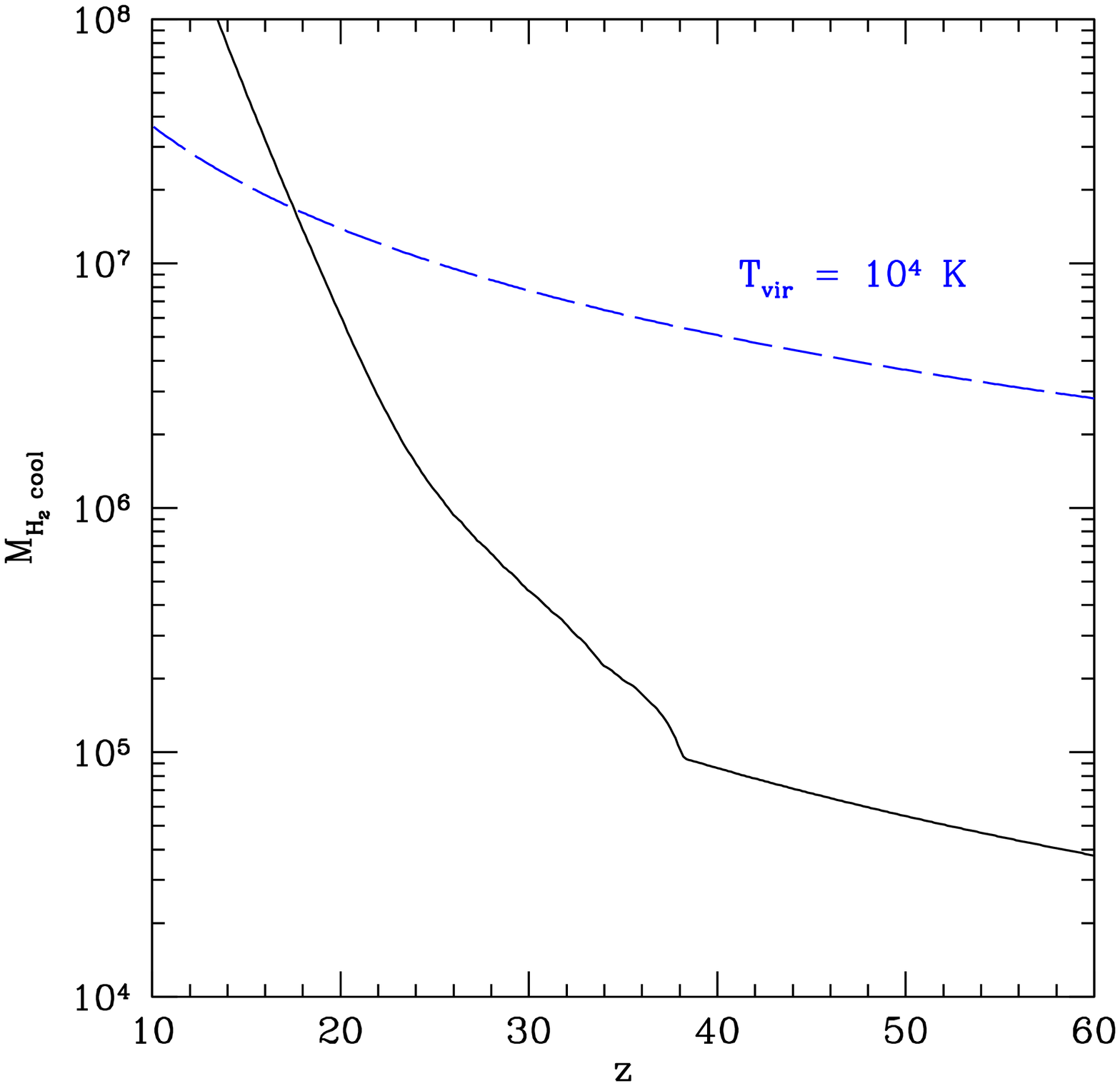}
  \plottwo{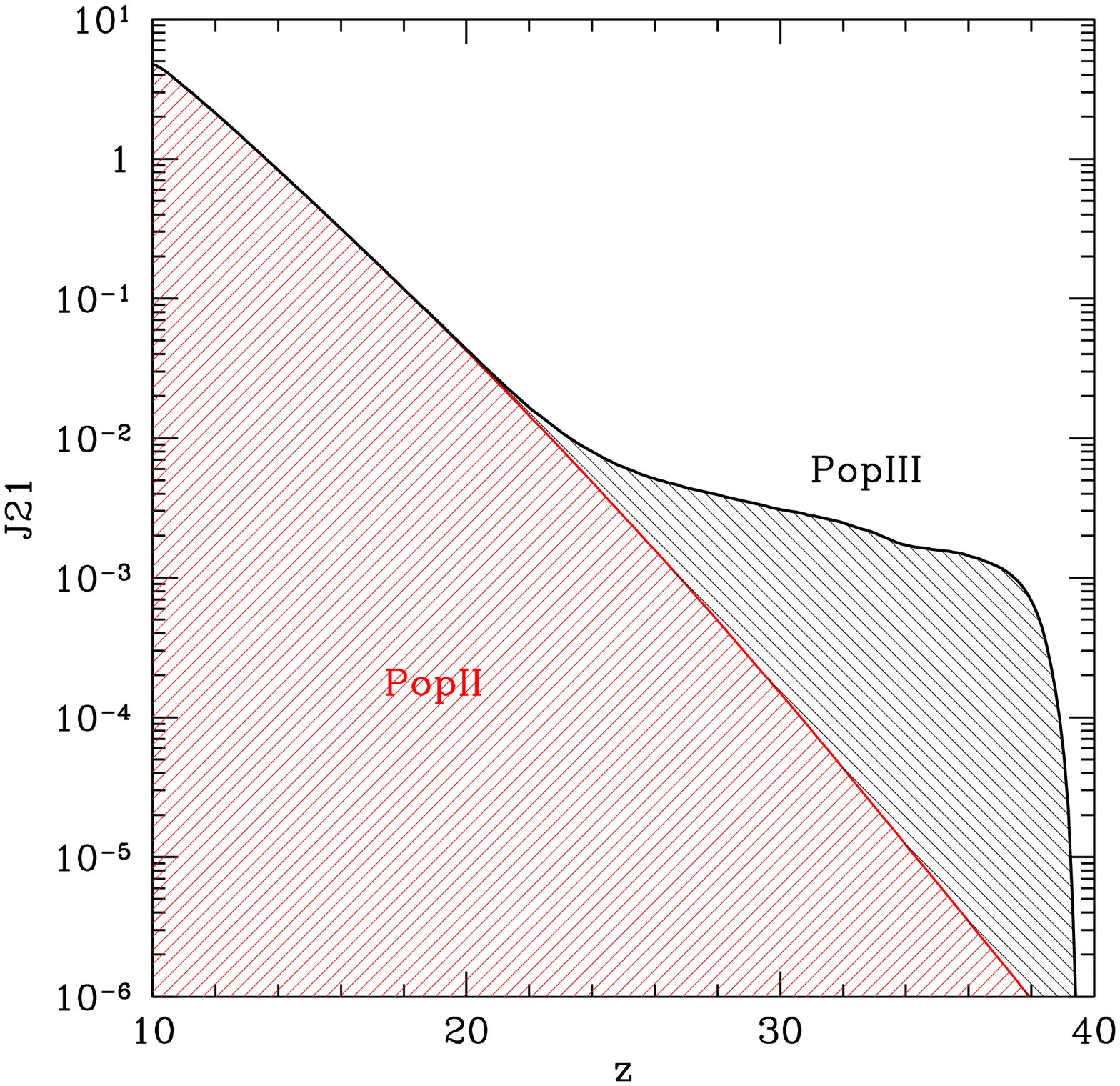}{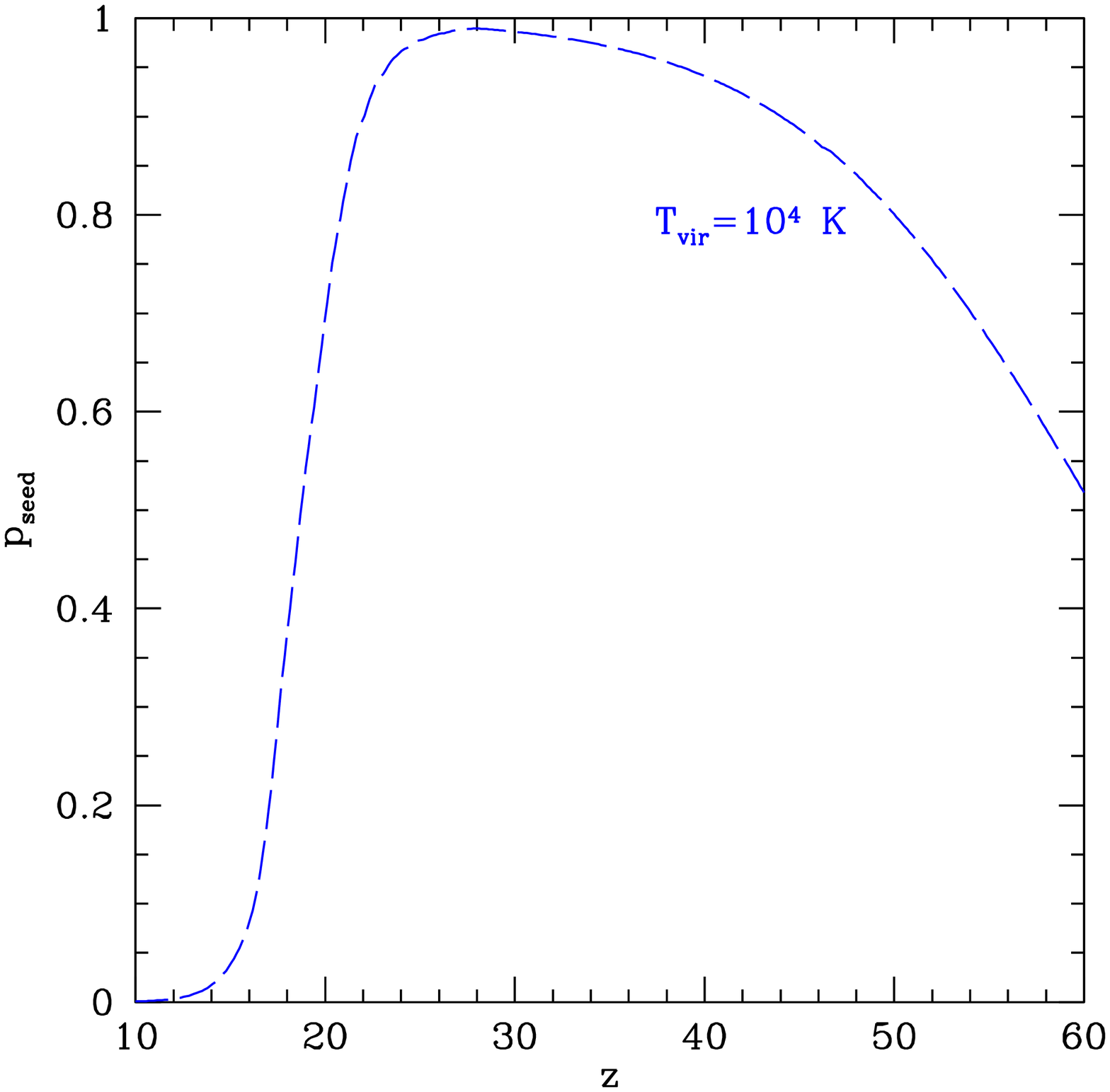}
\caption{Like in Fig.~\ref{fig:popIII_sfr_std} but for our model with
a strong external $J_{21}$ radiative field (see
eq.~\ref{eq:ext_LW}. This model yields similar results to our standard
model in Fig.~\ref{fig:popIII_sfr_std} because once the LW feedback is
above the critical threshold necessary to quench star formation in
minihalos its further increase has a modest additional effect.}\label{fig:extJ21}
\end{figure}

The qualitative picture described in our standard model holds even if
we vary the free parameters that regulates radiative feedback. In
Fig.~\ref{fig:low_escape} we show the star formation history when LW
feedback is strongly suppressed. Population III stars are formed at a
higher rate at $z \lesssim 30$ compared to the standard model, but the
difference is only a factor of a few (here the formation rate goes up
to $\approx 5 \times 10^{-5} \mathrm{M_{\sun} Mpc^{-3} yr^{-1}}$. This
is about one fourth of the peak formation rate without any radiative
feedback (see Fig.~\ref{fig:no_feedback}). The main difference from
the standard scenario is that Population III stars formed in halos
with $T_{vir} \geq 10^4 K$ are now suppressed even at $z \lesssim 15$,
because the enrichment probability of massive halos remains high (see
bottom right panel in fig.~\ref{fig:low_escape}). The formation of
Population II stars is instead essentially as in the standard
model. Note that the picture does not deviate significantly from the
standard model if we vary the efficiency of star formation in
Population II stars, except of course for a corresponding proportional
variation of their star formation rate. In fact Population III stars
in minihalos are the main agents of the radiative feedback that leads
to their suppression at $z \gtrsim 20$ (see bottom left panel in
Figs.~\ref{fig:popIII_sfr_std}-\ref{fig:low_escape}). Therefore their
formation is self-regulated and tends to reach an equilibrium level.
Similar results hold even when we add a stronger radiative feedback
based on the fixed Lyman Werner background given by
Eq.~\ref{eq:ext_LW} (see Fig.~\ref{fig:extJ21}). Note that in this
case the background radiation greatly exceeds that created by
Population III stars at $z \lesssim 20$ and thus the suppression of
$H_2$ cooling is even sharper.

\begin{figure} 
  \plottwo{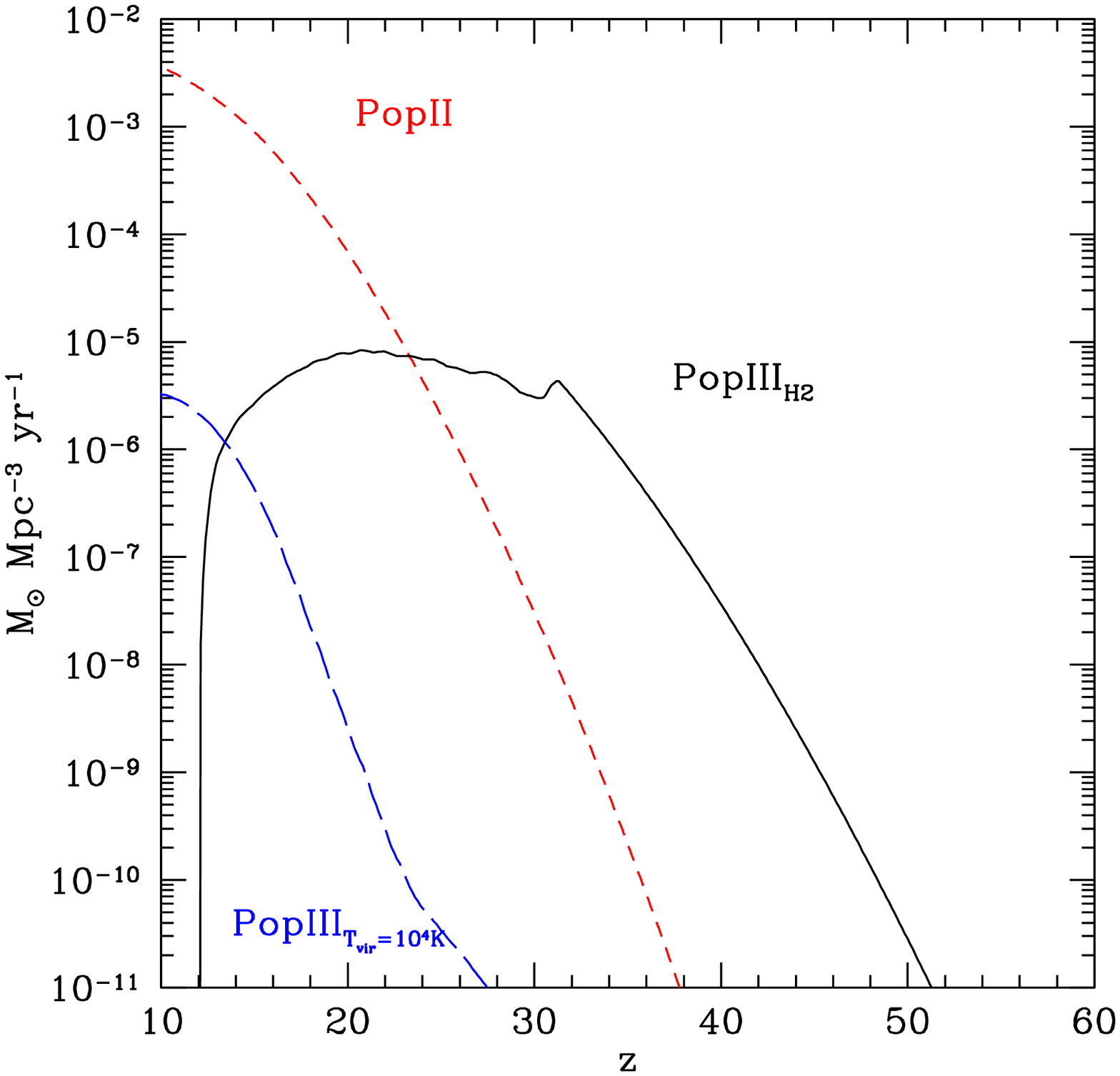}{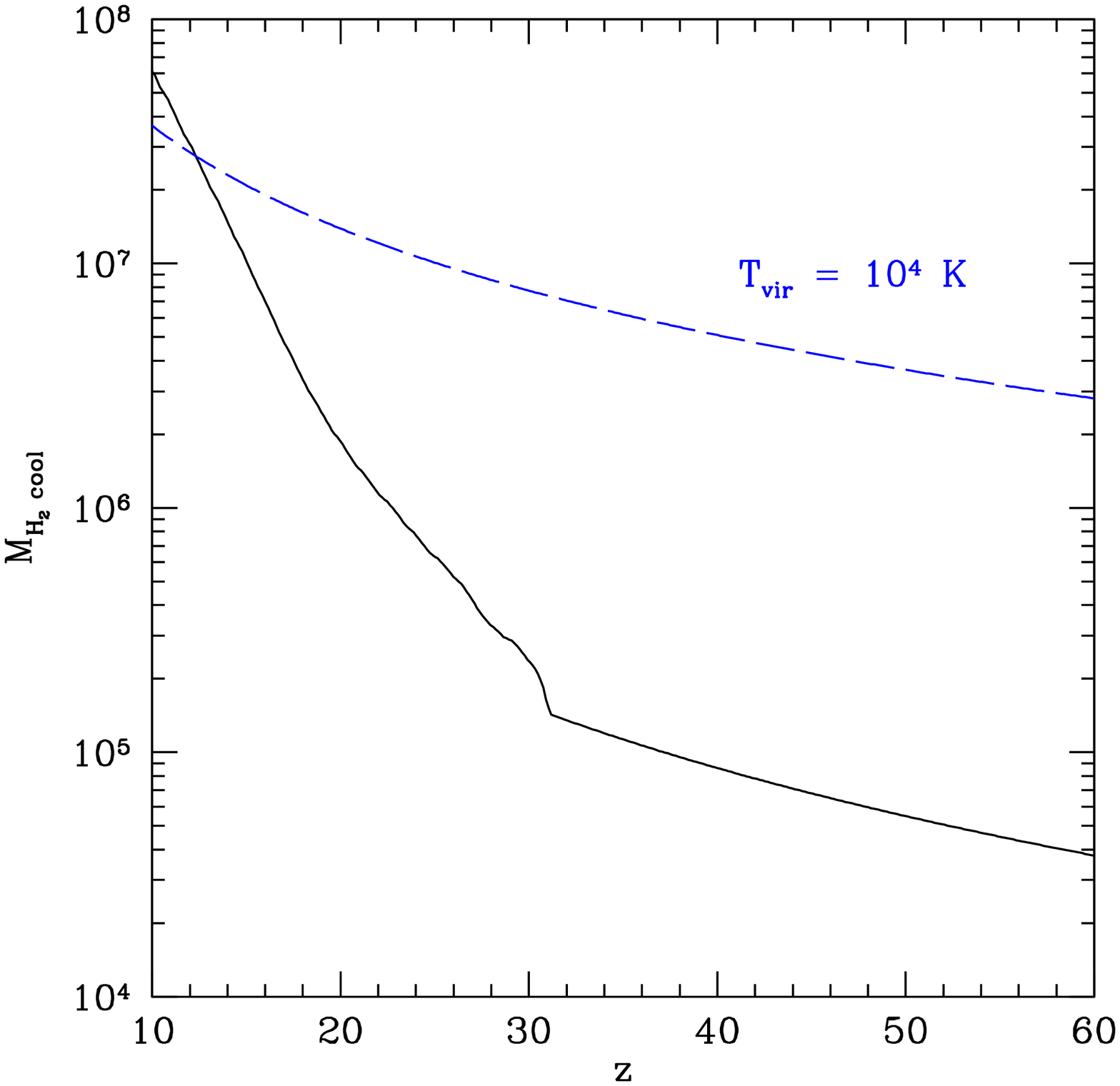}
  \plottwo{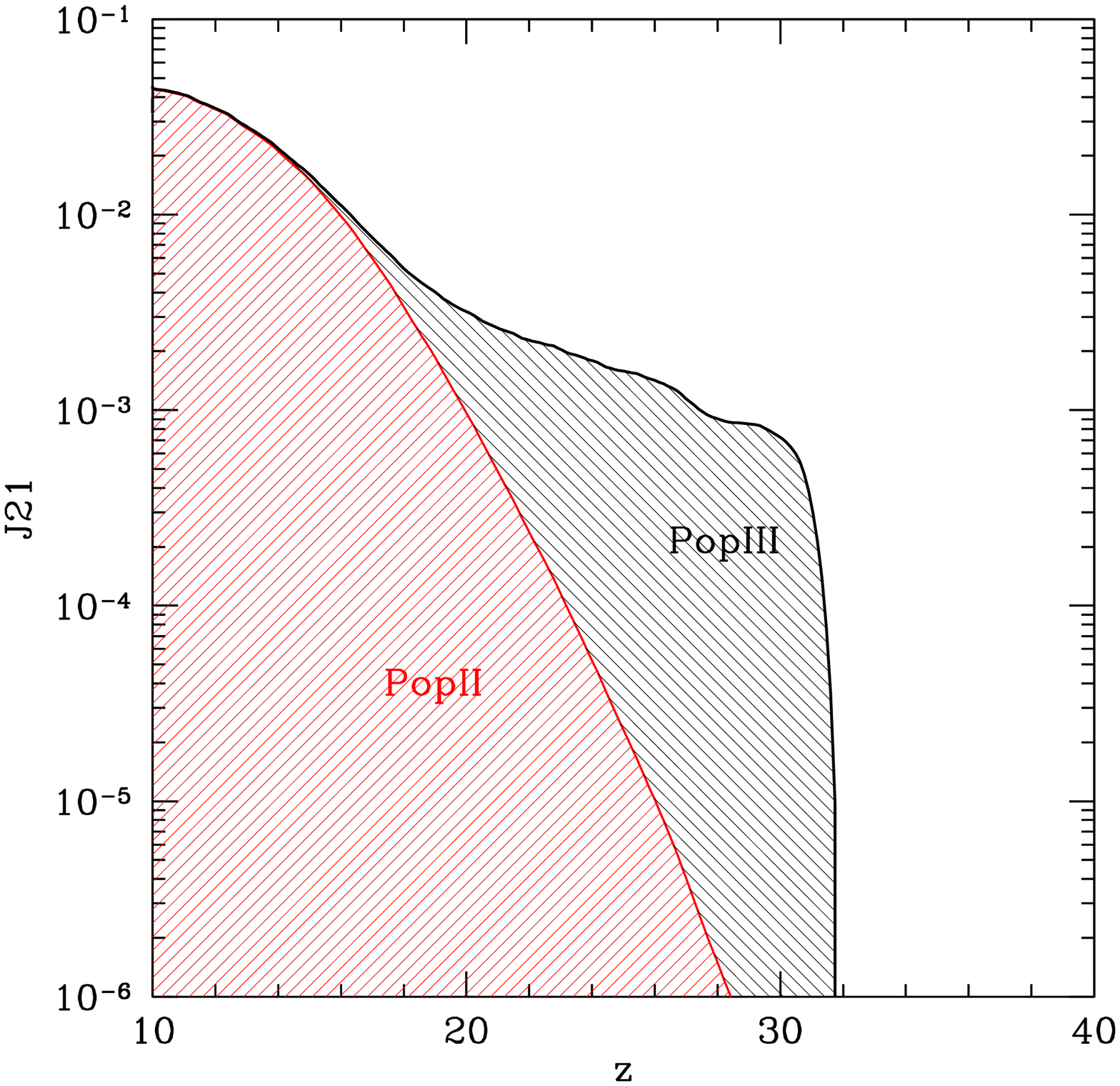}{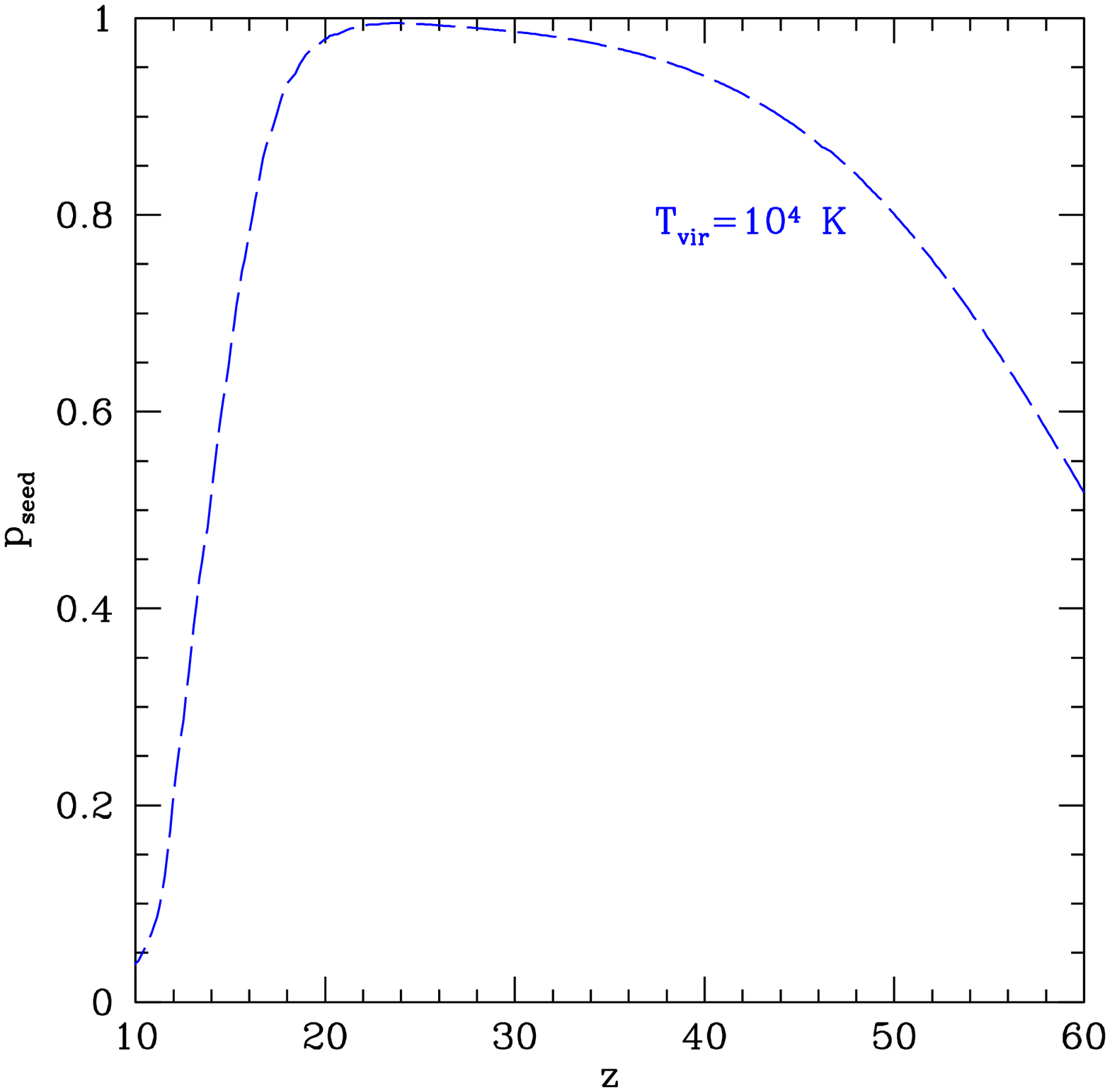}
\caption{Like in Fig.~\ref{fig:popIII_sfr_std} but for our model where
the halo mass function is computed using the \citet{PS} formula. The
star formation is suppressed at very high redshift, but by $z\approx
30$ the model closely resembles our standard model where the
\citet{she99} mass function is used. In fact, negative feedback acts
as a self-regulator of the Population III star formation
rate.}\label{fig:ps}
\end{figure}

Fig.~\ref{fig:ps} shows a model with our standard parameters except
for the use of the \citet{PS} mass function rather than the
\citet{she99} formula. The predictions for the Population III
formation rate are very different at $z>40$, but once the
self-regulated feedback phase starts the two models converge
together. The strong difference at $z>40$ originates from the fact
that Population III stars are hosted in very rare peaks at such early
times: in the \citet{PS} formalism these halos have $\nu =
\delta^2_c/\sigma^2(M) \gtrsim 30$ and the ratio of the Sheth-Tormen
to Press-Schechter mass function is proportional to $\exp(-0.707
\nu)/\exp(-\nu)$ in the limit of very large $\nu$. An interesting open
question is the form of the mass function for such rare peaks, which
are expected to be progressively more spherical as they become rarer
\citep{bardeen}. Fortunately the difference in the star formation
rate does not propagate significantly below $z \lesssim 30$.

A major qualitative change in the star formation history during the
Dark Ages arises if we allow for multiple Population III stars in a
single halo (see Fig.~\ref{fig:multiples}). In this case the metal
free star formation rate is comparable to that of second generation of
stars (Population II) down to $z\approx 10$. From the star formation
rate we can identify two eras: an early period dominated by Population
III in minihalos up to $z \approx 18$ and a later period where larger
halos are still able to form metal free stars. This happens because by
allowing multiple Population III stars in minihalos their formation
rate is significantly enhanced at later redshift over the assumption
of a single star per halo. In fact, when a significant LW background
is present, a single minihalo can have up to $10^6 \mathrm{M_{\sun}}$
of gas, thus a $\epsilon_{PopIII} = 5 \times 10^{-3}$ corresponds to a
SFR 50 times higher than the one obtained for a single metal free star
per halo. In this scenario metal free stars formed in $T_{vir}=10^4 K$
halos appear to be several orders of magnitude more common than in our
standard scenario, again because the star formation efficiency is
increased by more than two orders of magnitude compared to our
standard model. They are therefore expected to dominate the production
of ionizing photons (see Sec.~\ref{sec:reion}) until their formation
is eventually expected to be terminated by chemical enrichment due to
winds at $z \lesssim 15$.

\section{Consequences for the chemical enrichment of $T_{vir}=10^4$ K halos}\label{sec:enrich}

The first generation of Population III stars --- formed in minihalos
--- releases metals into the IGM and opens the way to metal enriched
star formation when the gas recollapses later as part of a larger
halo. When the second generation halo has a virial temperature
$T_{vir} \approx 10^4 \mathrm{K}$, the metallicity of its gas is
expected to have been enriched to $Z \gtrsim 10^{-4} Z_{\sun}$, high
enough to mark a transition toward a different mode of star formation,
especially if some dust is present \citep{schneider06,clark08}. Our
model allows us to quantify the likelihood of this self-enrichment
scenario as a function of the redshift of formation of a
$T_{vir}=10^4$ K halo (see bottom right panel in
Fig.~\ref{fig:popIII_sfr_std}). Interestingly, the probability of
having a pristine halo large enough to reach $T_{vir}=10^4 K$ is small
at very high redshift ($z\approx 30$) and progressively increases as
the redshift decreases. This apparently surprising result can be
understood in terms of the difference in halo mass required to cool
via molecular and atomic hydrogen. At $z \gtrsim 30$, the difference
in the two masses is large (see upper right panel in
Fig.~\ref{fig:popIII_sfr_std}), hence it is highly likely that in the
merger tree of the $T_{vir}=10^4$ K halo there has been a progenitor
halo that was able to form a Population III star via $H_2$ cooling at
an earlier redshift. However, as the redshift decreases, the LW
self-shielding mass grows and it becomes progressively more unlikely
that one of the progenitor halos hosted a star. Therefore halos with
$T_{vir} \approx 10^4 \mathrm{K}$ are more likely to be metal free at
lower redshift. Our model does however not include metal pollution by
winds, which becomes progressively more important as the redshift
decreases. In fact, the wind pollution model by \citet{fur05} predicts
a sharp drop in the probability of forming pristine halos around $z
\approx 15$. Therefore if we combine our reference scenario results
with the \citet{fur05} model for the wind enrichment (e.g. see their
Fig.~2), the overall picture is that the pollution probability for
such halos remains high at all redshifts, probably presenting a
minimum (with enrichment probability down to $\approx 50\%$) between
$z=15$ and $z=20$. This further strengthens our conclusion that
Population III star formation in halos with $T_{vir} \geq 10^4 K$ is
expected to be subdominant compared to formation in minihalos if
primordial stars are formed in isolation. This conclusion becomes
stronger as the LW feedback efficiency is decreased or set to zero
(see figs.~\ref{fig:low_escape} and~\ref{fig:no_feedback} ).

A different possibility for star formation in massive metal free halos
is the creation of a quasi-star, that is a black hole formed via
direct collapse accreting inside a massive envelope
\citep{beg08}. This scenario has been proposed to explain the rapid
growth of supermassive black holes, but a critical requirement for its
viability is the absence of metals in the gas \citep{omu08}. Our
results on chemical enrichment suggests that quasi-stars are likely to
be formed only in a redshift window between $z \approx 20$ and $z
\approx 10$. At the lower end of this redshift window quasistars might
only live in regions that lack primordial galaxies and thus have a
relative suppression of structure formation compared to the average
over the whole Universe. The environment where the progenitors of
bright high redshift quasars live is unlikely to qualify as one of
such regions. If we assume that a bright, rare, $z=6$ QSO is located
at the center of a dark matter halo of mass $M = 5 \times 10^{12}
\mathrm{M_{\sun}}$ \citep{MR05}, then we derive that its \emph{first}
progenitor with $T_{vir} > 10^4 \mathrm{K}$ has been formed at $z>26$
with a confidence level greater than $0.9999$. Such progenitor halo
has a probability of being metal enriched greater than $98\%$, so it
only has a small chance to host a quasi-star. Note however that
quasistars can still lead to the formation of a bright $z\approx 6$
QSO if they are first formed in a relatively void region at $z \approx
15$ and then merge by $z \approx 10$ into the gas rich environment of
the main QSO progenitor halo. This scenario allows the black hole seed
from the quasistar to grow with maximal efficiency. In fact, the BH
starts with a large mass from the initial direct collapse and then it
enters a gas rich region where it can grow continuously at near
Eddington limit. As a $z=6$ bright QSO halo consists of material
originating in a sphere with comoving radius larger than $3
\mathrm{Mpc}$, this is not unlikely. In fact a wind traveling at
speeds of $30 \mathrm{km/s}$ only covers a fraction of this distance
in half a billion years. However a detailed modeling can be obtained
only through the study of QSO merger trees that include information on
the spatial distribution of the progenitors, such as those given by
cosmological simulations.
 
\section{Population III stars and the Reionization of the Universe}\label{sec:reion}

The role of Population III stars in the reionization of the Universe
has been much debated in the last years, especially after the first
year WMAP data release, which included a high optical depth to
reionization $\tau_e = 0.17 \pm 0.03$ \citep{spe03}. Several models
had been proposed \citep{ven03,cen03,wyi03,hui03,sti05} and many of
these included a significant contribution from first stars to
$\tau_e$. With the latest WMAP data release, the optical depth to
reionization is rather low ($\tau_e = 0.084 \pm 0.016$ implying an
instantaneous reionization redshift $z_{reion} \approx 10.0 $; see
\citealt{kom08}) and its major contribution comes from complete
ionization after $z=6$ \citep{shu08}. The contribution from higher
redshift is limited to $\Delta \tau_e = 0.03 \pm 0.02$, providing an
upper limit to the luminosity of primordial galaxies \citep{shu08}.

From our study it appears that the contribution of Population III
stars to the total budget of reionizing photons is limited if only one
star per halo is formed, even neglecting negative radiative
feedback. Despite the fact that first stars are more than one order of
magnitude more efficient at producing ionizing photons per unit mass
than Population II stars \citep{tum00,schae03} their overall star
formation rate in our standard model is significantly lower for $z
\lesssim 20$. Based on our standard model and assuming a Population
III formation rate of $10^{-5} \mathrm{M_{\sun} Mpc^{-3} yr^{-1}}$
from $z=35$ to $z=10$, we obtain that about $n \approx 4 \times
10^{65} \mathrm{Mpc^{-3}}$ ionizing photons are emitted by metal free
stars. This falls short of the number density of hydrogen atoms $n_{H}
\approx 7 \times 10^{66} \mathrm{Mpc^{-3}}$. Thus after taking into
account the effect of clumpiness of the IGM and of recombination, it
is clear that Population III stars can only reionize a minor fraction
of the hydrogen atoms, even if the escape fraction is near unity. A
large escape fraction is in fact possible in minihalos ($M \lesssim
10^6 \mathrm{M_{\sun}}$), but is likely significantly smaller in
larger halos ($M \lesssim 10^6 \mathrm{M_{\sun}}$), where the H II
region may remain confined well within the virial radius of the host
halo \citep[see][]{whalen04,kita04}. The number of ionizing photons
produced is smaller than the number of hydrogen atoms even in our
model with no negative feedback (see Fig.~\ref{fig:no_feedback}). In
order for Population III stars to be a significant agent of
reionization multiple PopIII stars must be formed in a single
halo. For our model with $\epsilon_{PopIII} = 0.005$, we obtain a
cumulative ionizing photon production of $\approx 4 \times 10^{67}$
down to $z=10$. Such number of photons starts to become sufficient to
contribute to reionization even for a relatively low escape fraction
($f_{esc} \sim 0.1$). Certainly Population III stars are major agents
of reionization if their star formation efficiency goes up one order
of magnitude to $\epsilon_{PopIII}=0.05$ (see
Fig.~\ref{fig:multiple_high_eff}). Note that in both these scenarios
with multiple metal free stars per halo, the main sources of
reionization are primordial galaxies in halos with $T_{vir}> 10^{4}
\mathrm{K}$. In fact, the reionizing photon budget from Population III
remains significant even when only one star per minihalo is formed,
but clusters of metal free stars are allowed in larger halos (see
Fig.~\ref{fig:hybrid}; see also \citealt{haiman06}).

\section{Conclusions}\label{sec:conc}

In this paper we present a model for the star formation rate of
metal free (Population III) and second generation (Population II)
stars during the Dark Ages of the Universe, at $z \geq 10$. The model
relies on dark matter halo mass function coupled with analytical
prescription for cooling and collapse of gas clouds. Our model
includes radiative Lyman Werner feedback, which can suppress star
formation in minihalos, and self-enrichment feedback, which marks the
transition from metal free to Population II stars. 

Thanks to our novel treatment of chemical enrichment, based on the
formalism developed in \citet{ts07a}, we show that halos with a virial
temperature $T_{vir}\geq 10^4 \mathrm{K}$ are most likely to host a
second generation of stars, formed from gas enriched to a metallicity
$Z \geq 10^{-4} Z_{\sun}$ by a progenitor Population III star in a
minihalo at a higher redshift. Metal free stars can form in halos
with $T_{vir}\geq 10^4 \mathrm{K}$ only once the cooling of gas
in minihalos is strongly suppressed by radiative Lyman Werner
feedback, which in our reference model happens at $z \lesssim 20$. If
only one Population III star forms per dark matter halo, then their
number is dominated by those formed in minihalos with a peak star
formation rate of $\approx 10^{-5} \mathrm{M_{\sun} Mpc^{-3} yr^{-1}}$
at $z\approx 20$. This prediction is robust and does not depend on the
detail of the model. In fact, the negative radiative Lyman Werner
feedback acts as a self-regulator of star formation in minihalos
keeping variations of the star formation rate limited when the
feedback efficiency or the halo mass function is changed.

The metal enrichment also leads to an early rise of the star formation
rate of Population II stars. By redshift $z \lesssim 26$ their SFR is
higher than that of Population III stars and steadily rises as the
redshift decreases. In our model we do not include positive radiative
feedback that can promote $H_2$ formation in the neighborhood of a
first star \citep{ric01}. If this is the case, then the transition to
metal enriched stars in halos with $T_{vir}\geq 10^4 \mathrm{K}$ is
expected to be even more solid, because multiple Population III stars
in clustered minihalos can pollute to a higher metallicity the gas
that later constitutes a $T_{vir}\geq 10^4 \mathrm{K}$ halo. The metal
enrichment probability from minihalo pollution decreases once a strong
LW background is in place, so at redshift $z \lesssim 15$ halos
containing pristine gas with a mass $M \gtrsim 2 \times 10^{7}
\mathrm{M_{\sun}}$ are possible, provided that winds from
protogalaxies, absent in our model, are not too efficient in polluting
the IGM. Redshift $z\approx 15$ might thus be the most favorable
period for the formation of quasi-stars \citep{beg08}, which are
however expected to reside preferentially in underdense environments,
where winds are more unlikely to be present and pollute the IGM.

For this standard scenario the contribution to reionization given by
Population III stars is only indirect (they enrich the IGM and allow
Population II to form). In fact the cumulative number of ionizing
photons they produce falls shorter of the number of hydrogen
atoms. Metal-enriched stars in the first galaxies are thus expected to
be the main agents of reionization, even though the rapid decrease of
the galaxy luminosity function at $z>6$ \citep{bou06,oesch08} casts
some doubts on this scenario.

\begin{figure} 
  \plottwo{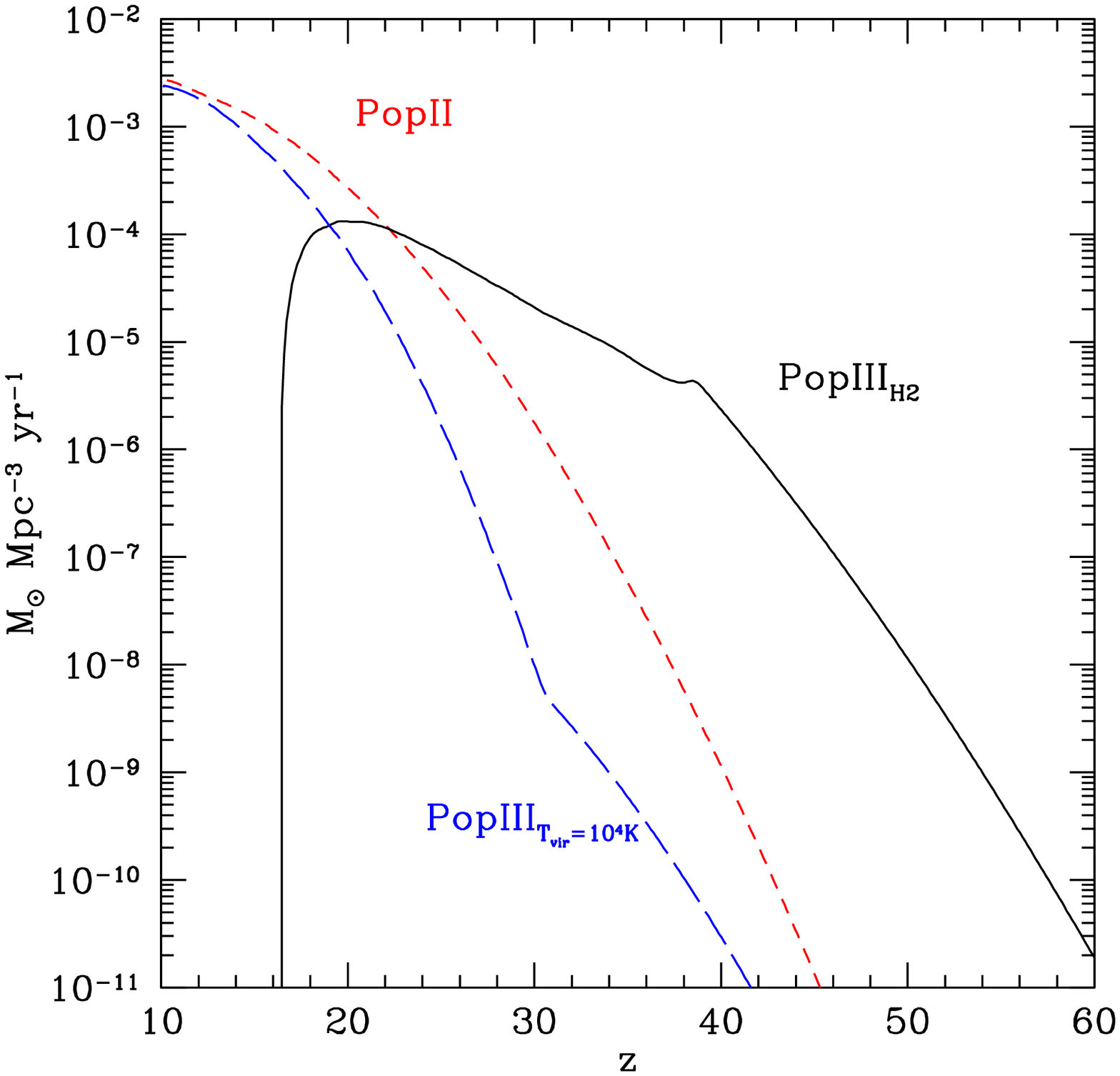}{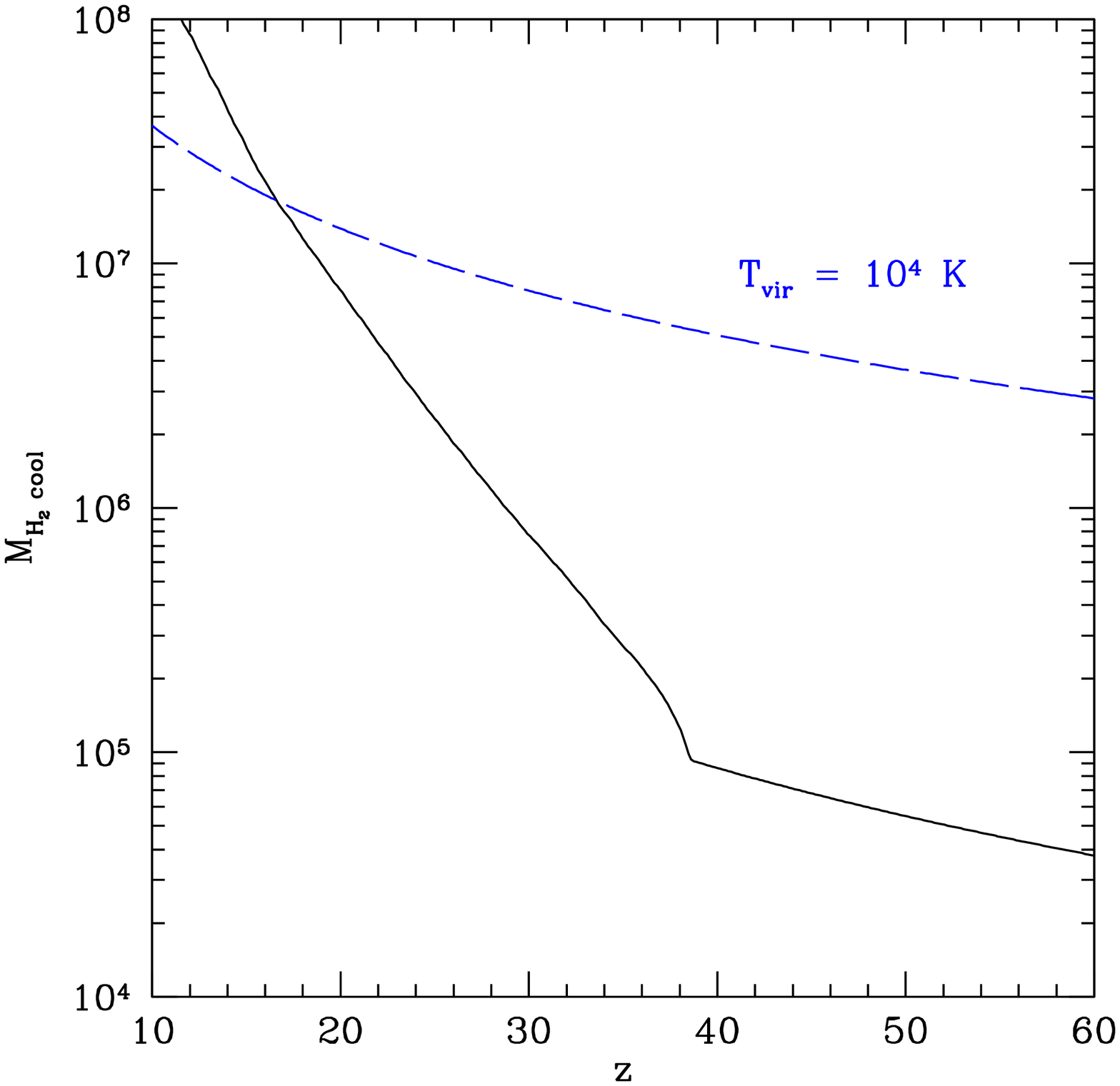}
  \plottwo{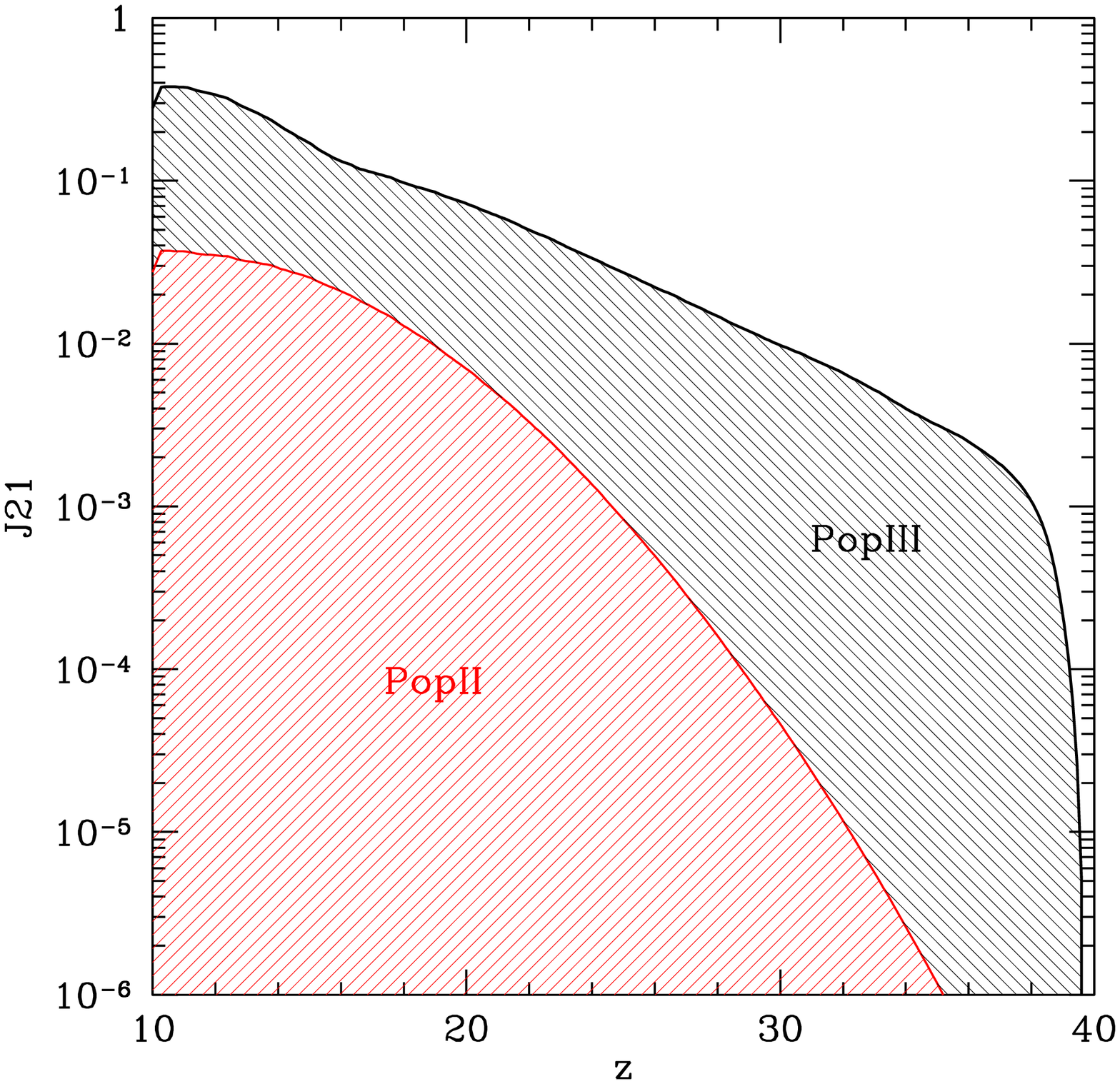}{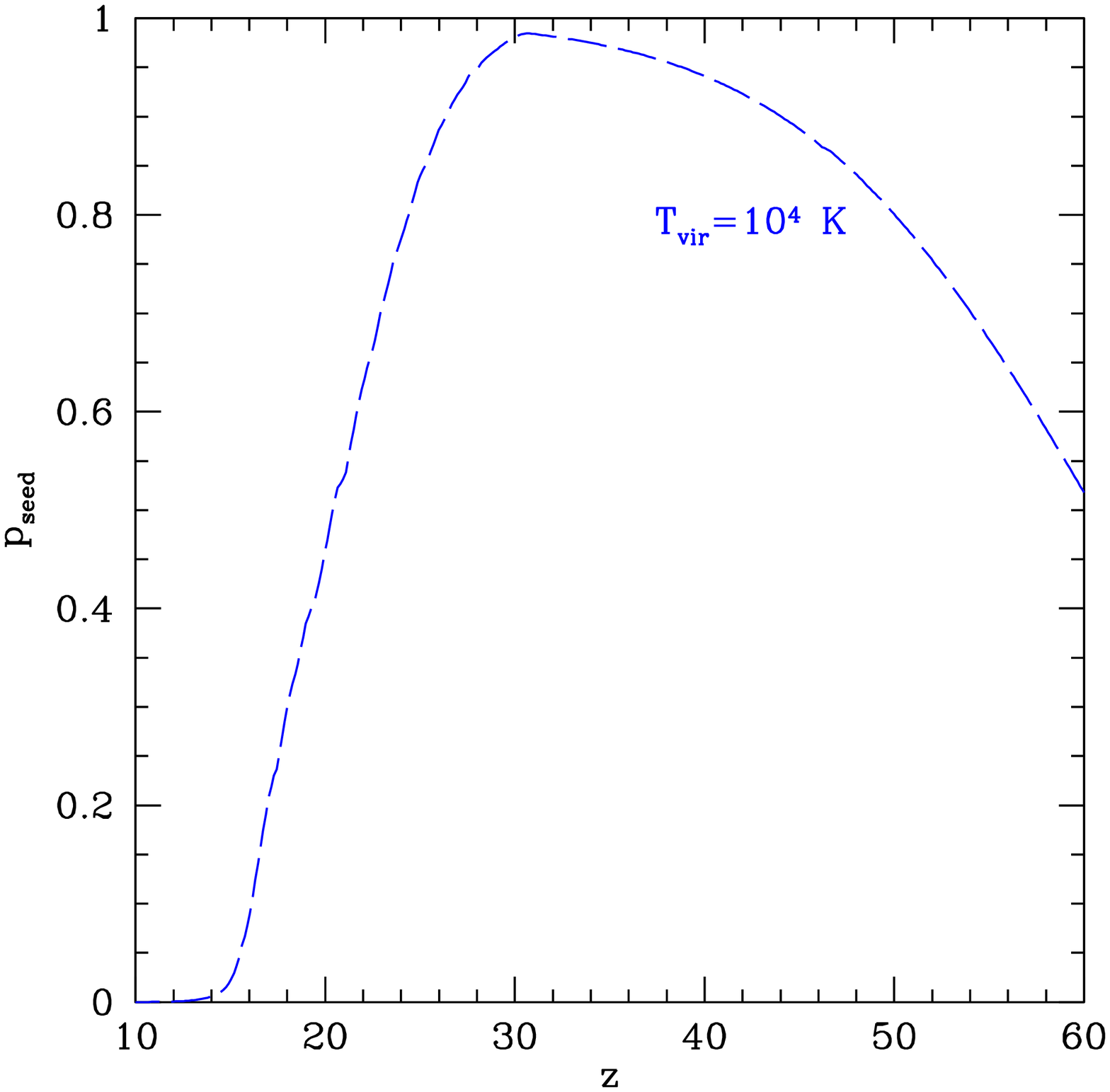}
\caption{Like in Fig.~\ref{fig:popIII_sfr_std} but for our model with
multiple Population III stars allowed to form in the same halo (with
efficiency $\epsilon_{PopIII}=0.005$).}\label{fig:multiples}
\end{figure}

\begin{figure} 
  \plottwo{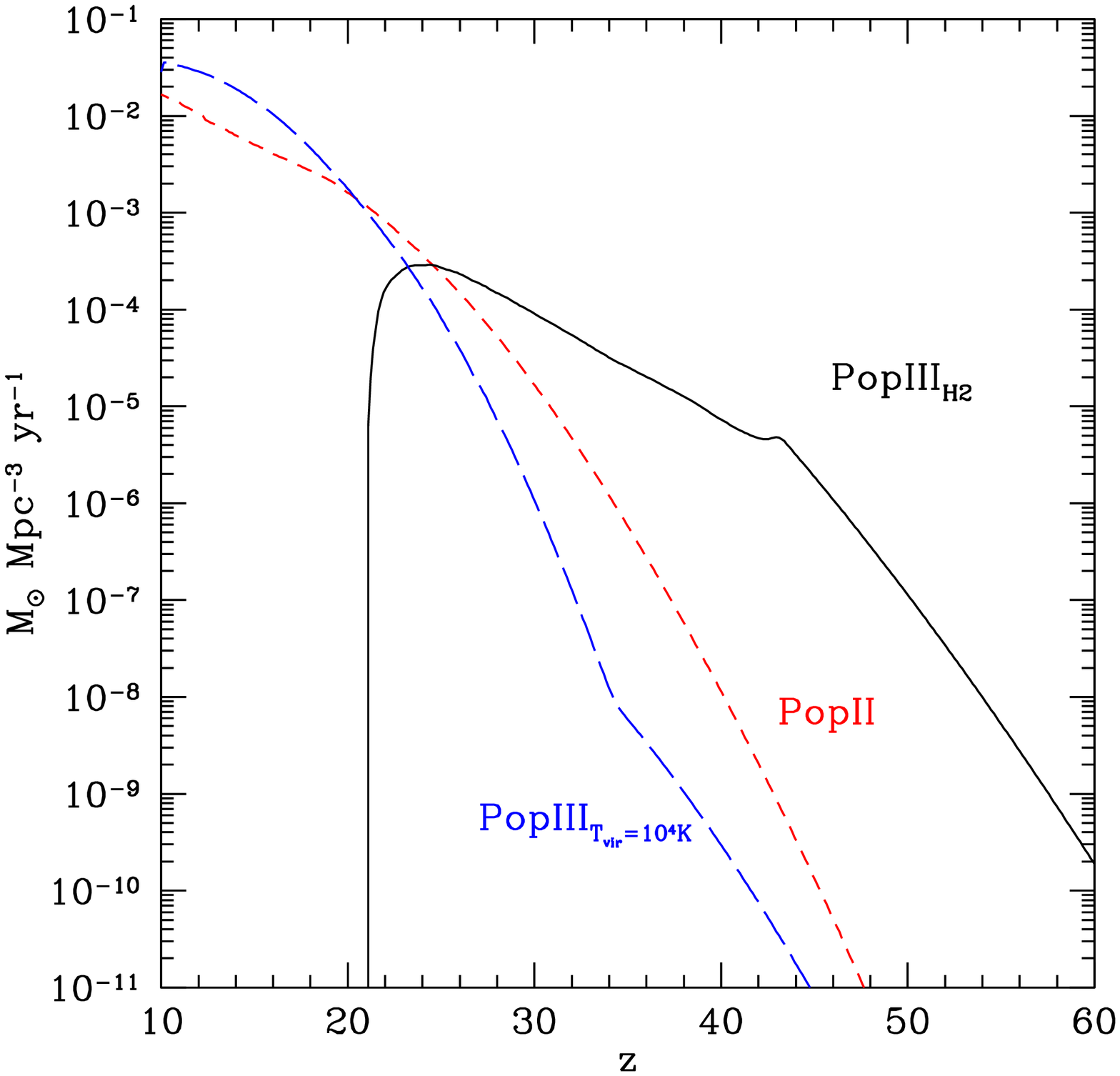}{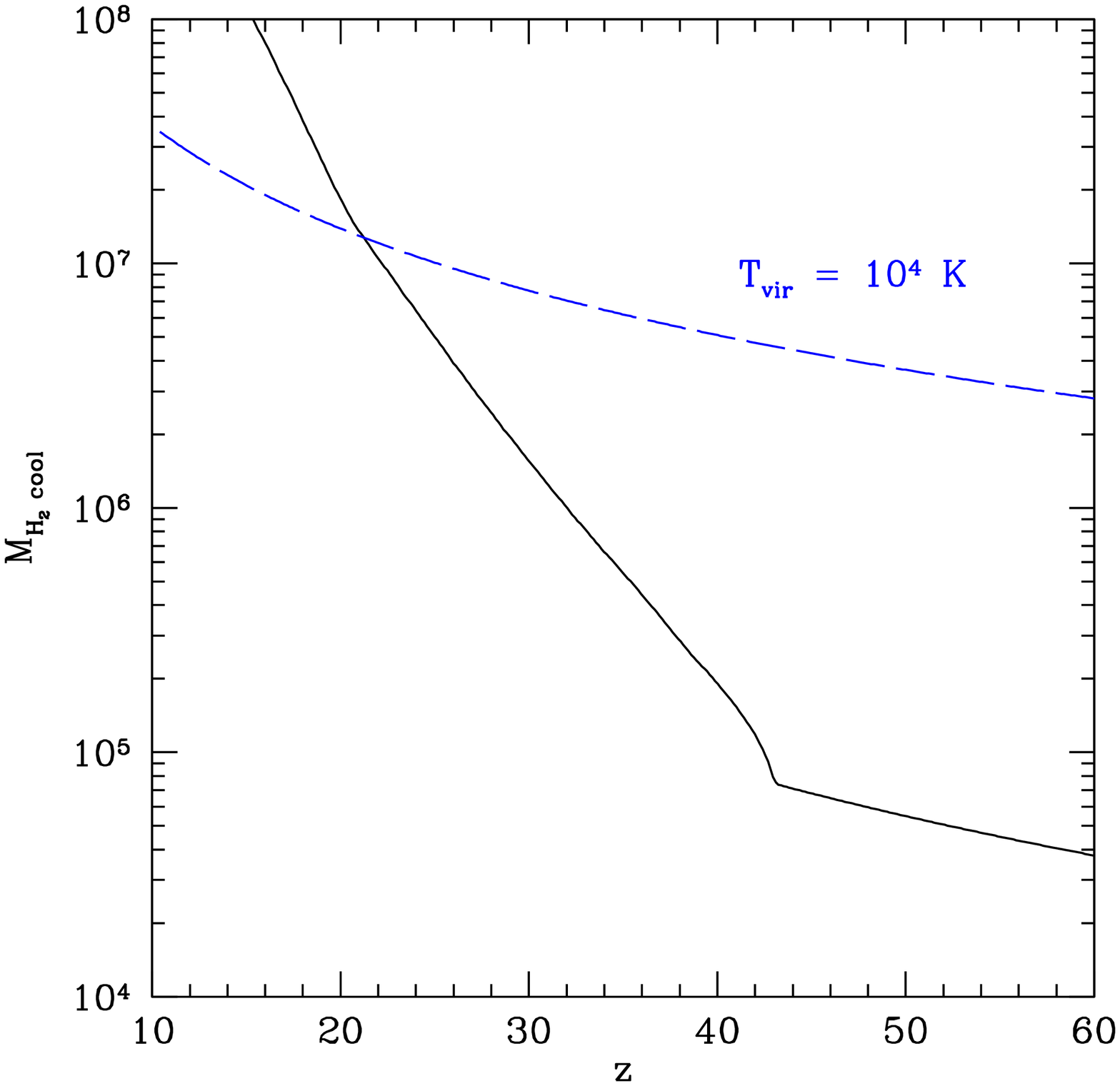}
  \plottwo{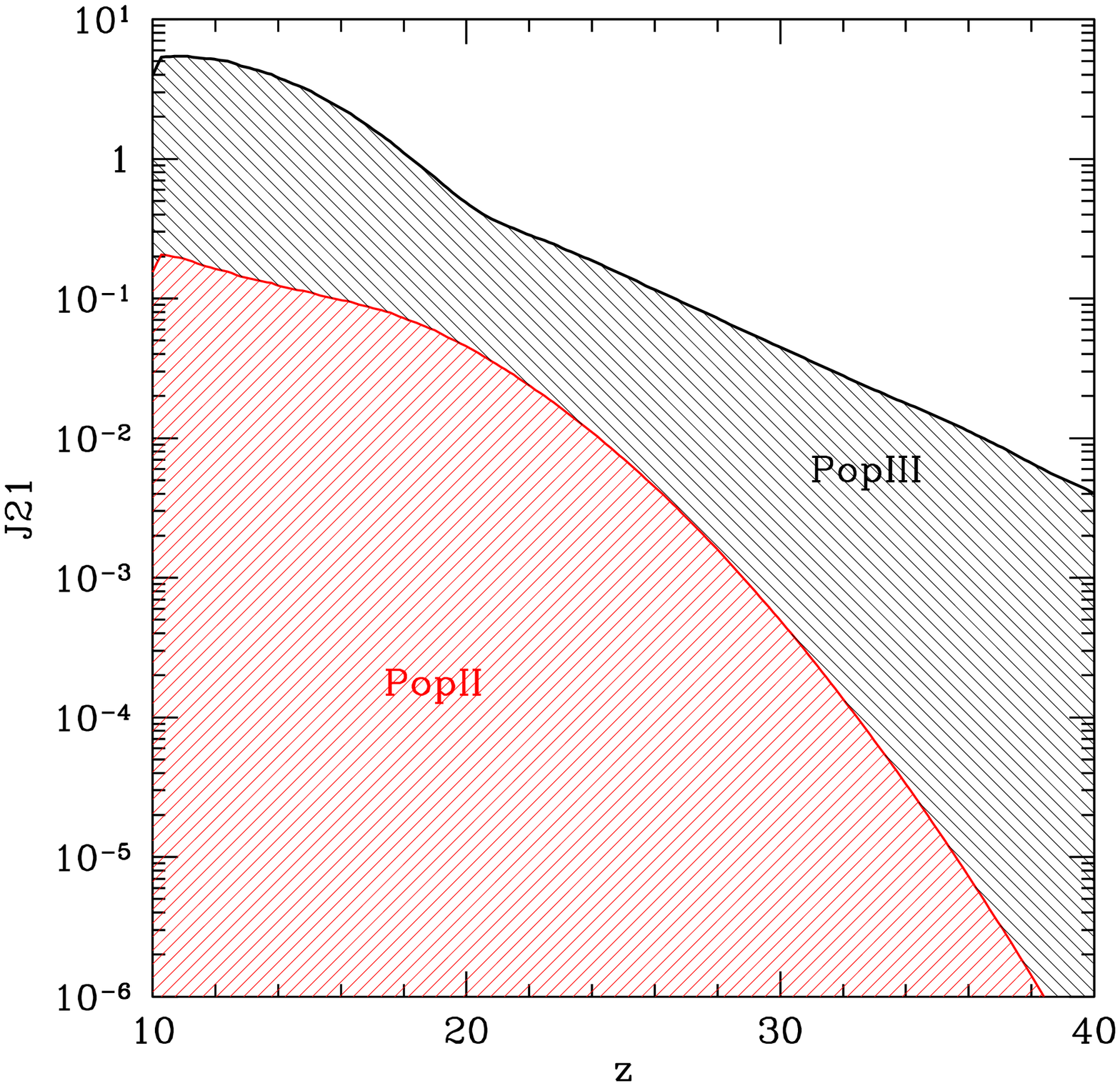}{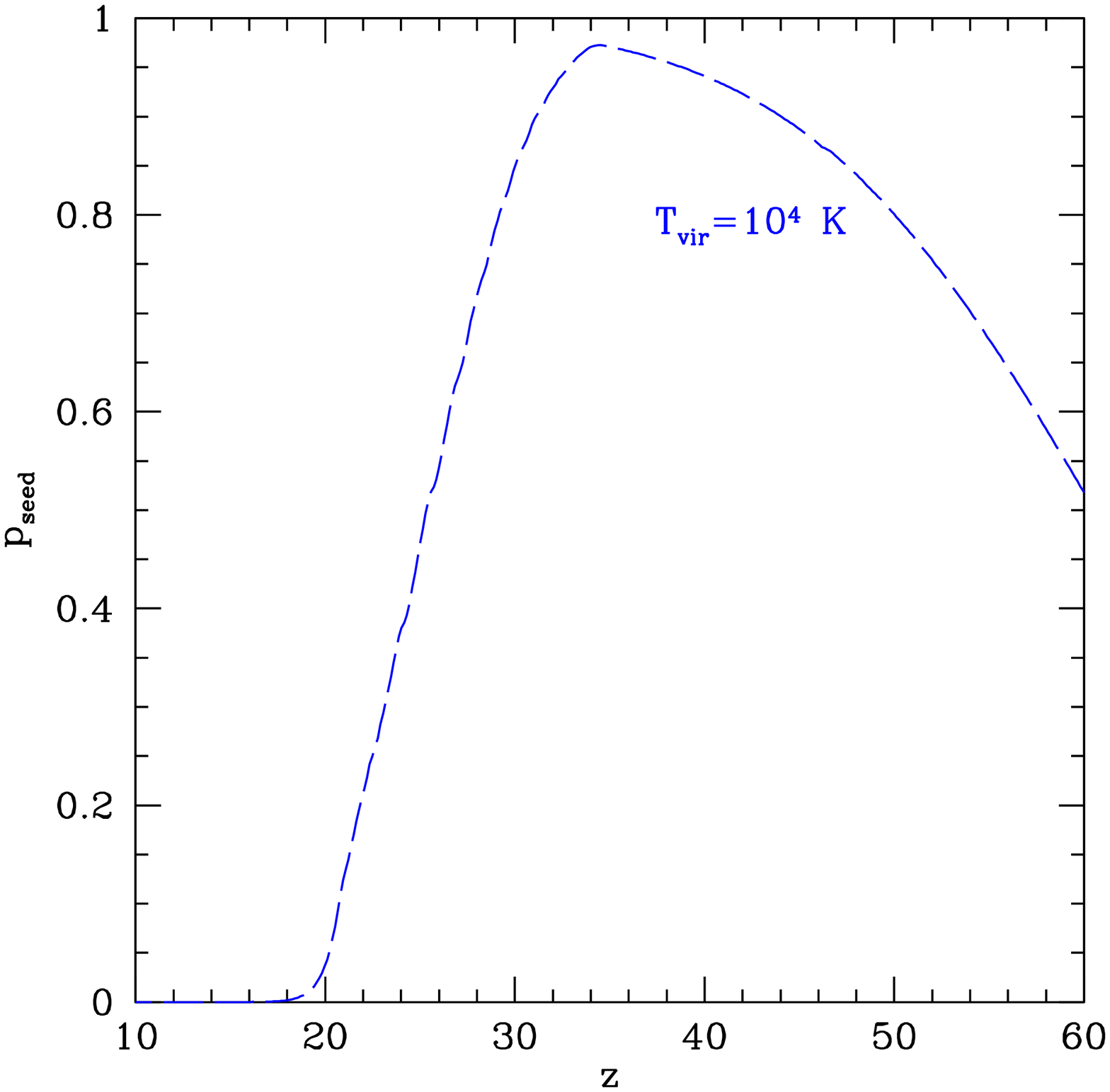}
\caption{Like in Fig.~\ref{fig:popIII_sfr_std} but for our model with
multiple Population III stars allowed to form in the same halo and
have an enhanced efficiency
($\epsilon_{PopIII}=0.05$).}\label{fig:multiple_high_eff}
\end{figure}

\begin{figure} 
  \plottwo{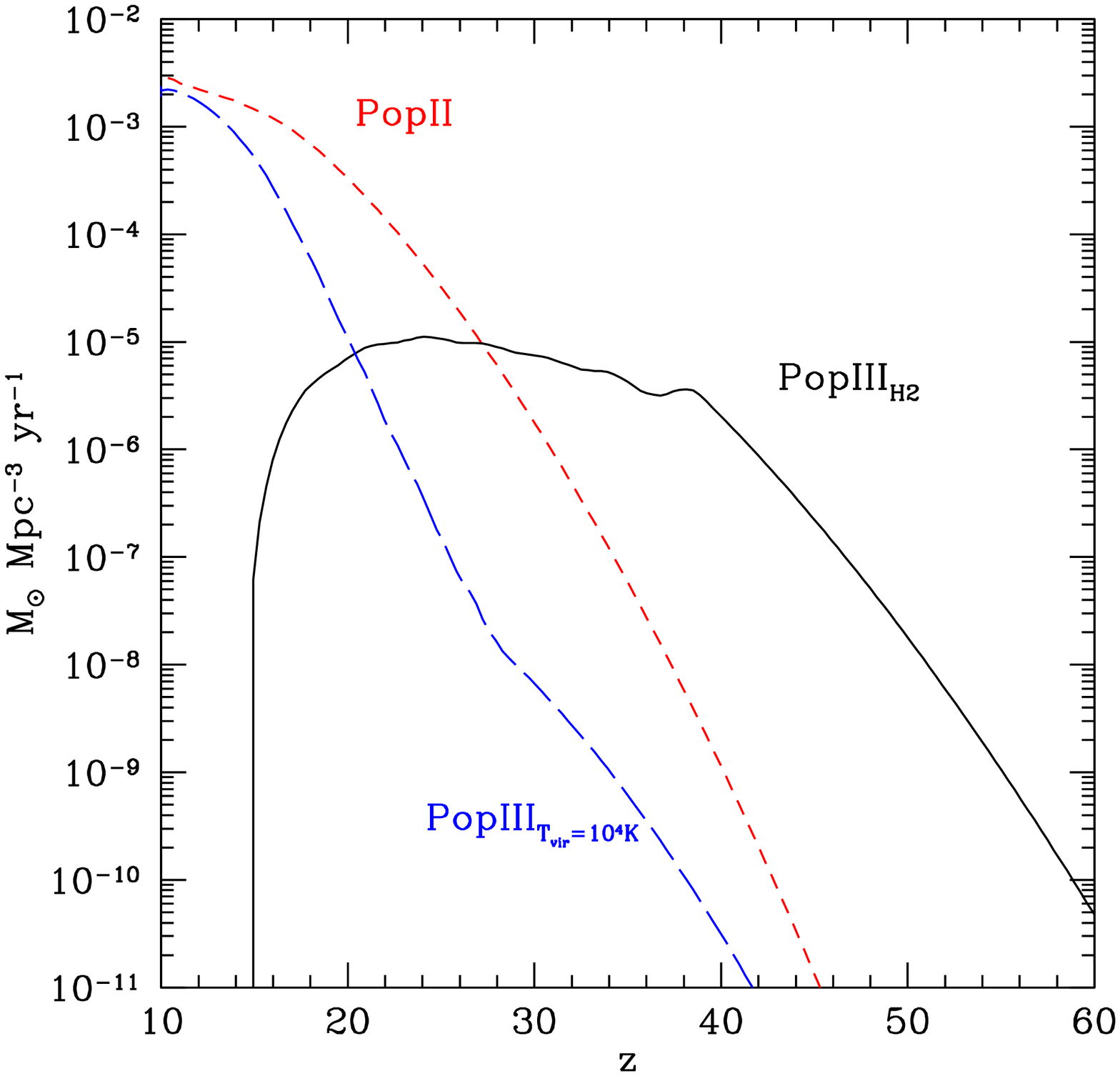}{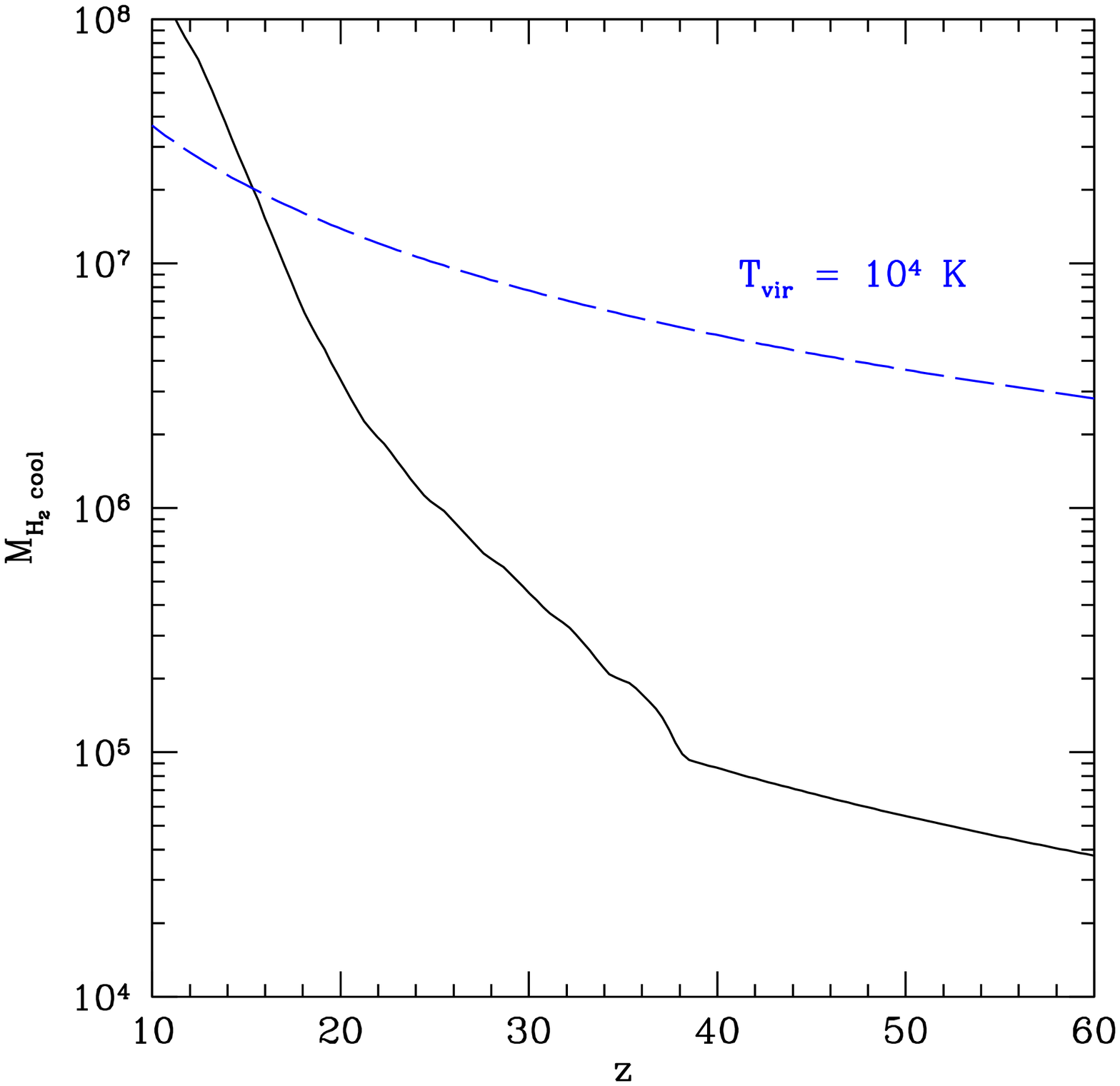}
  \plottwo{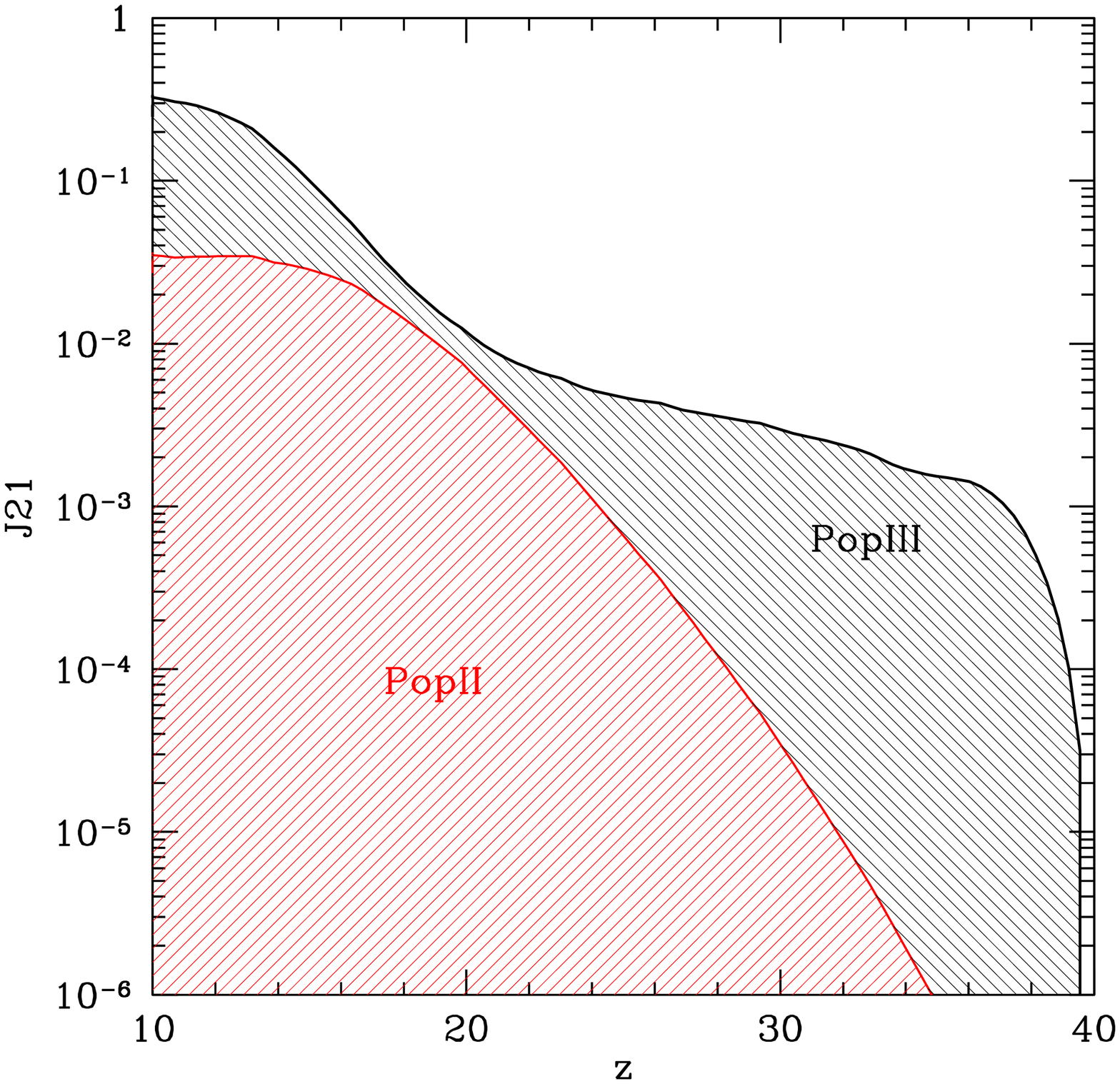}{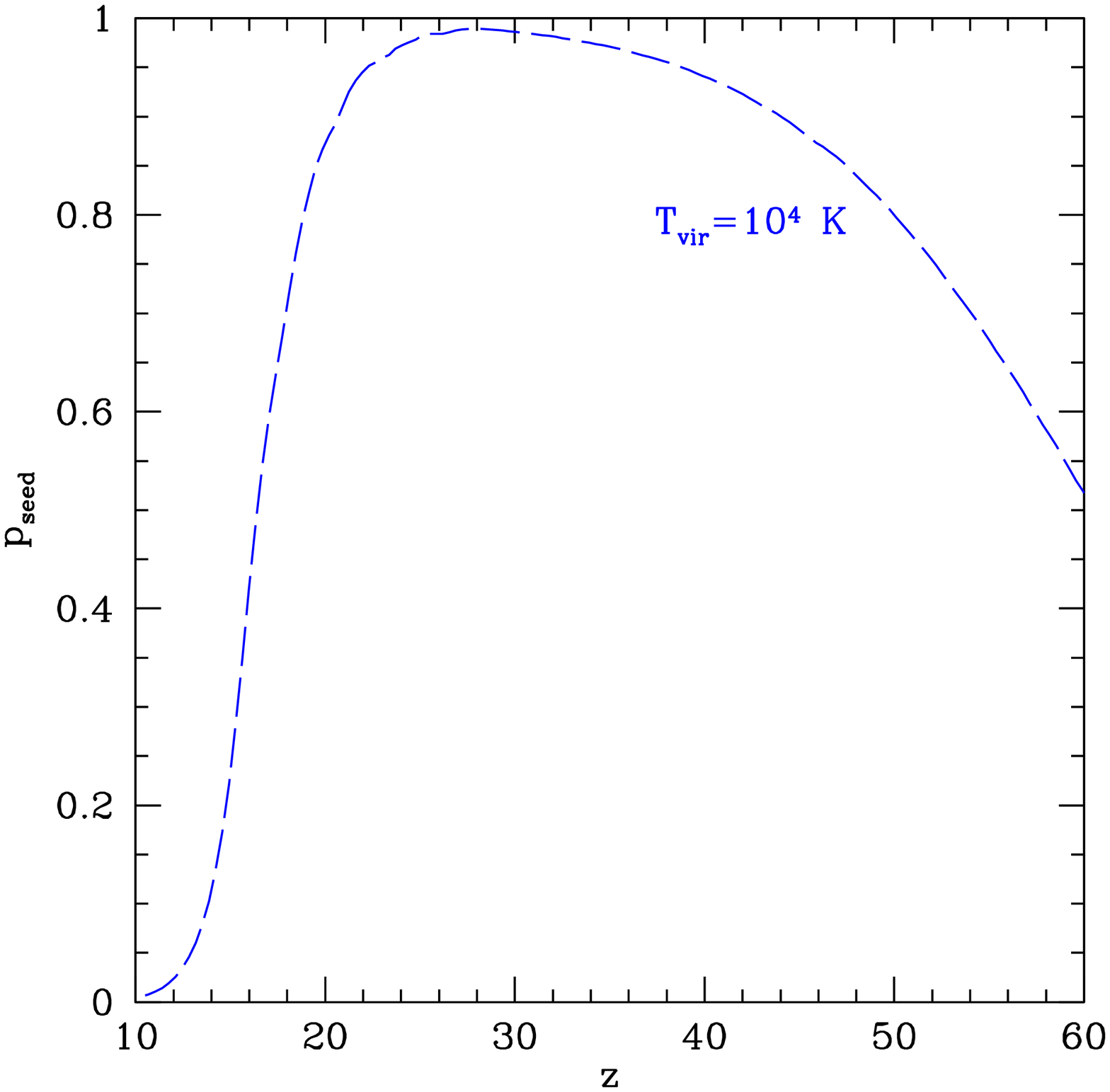}
\caption{Like in Fig.~\ref{fig:popIII_sfr_std} but for our model with
multiple Population III stars allowed to form in halos with $T_{vir}
\geq 10^4 \mathrm{K}$ (with efficiency
$\epsilon_{PopIII\_T_{vir}=10^4 K}=0.005$) and a single star per
minihalo.}\label{fig:hybrid}
\end{figure}

A main change in the Population III star formation rate, which
increases their contribution to reionization, can be introduced if one
releases the assumption that only a single metal free star is formed
per halo: by converting a fixed fraction of gas into primordial stars
in minihalos the feedback mechanism is less efficient because as the
critical gas mass needed for cooling is increases so is the number of
stars produced per halo. Therefore in this scenario there is a
constant growth of the Population III star formation rate in minihalos
until this formation channel is suddenly inhibited because the minimum
mass required to self-shield molecular hydrogen in the halo
corresponds to a virial temperature above $10^4$ K (see
fig.~\ref{fig:multiples}). Under this scenario Population III stars can
easily produce a significant amount of ionizing photons and could
account for a significant fraction of the optical depth to
reionization originating from $z>7$. One issue to be addressed if
multiple Population III stars are formed in a single halo is however
the impact of local radiative feedback. The first massive star formed
in a minihalo emits enough energy to completely photo-dissociate the
gas in the halo, thus multiple stars can be formed only if there is a
single star formation burst of limited time duration and very high
efficiency. This is a crucial issue that can be properly addressed
only once cosmological simulations of Population III star formation
will be able to go past the formation of the first protostellar core
and follow multiple episodes of star formation (see \citealt{wise08}
for promising progress in this direction).

\acknowledgements

This work was supported in part by NASA JWST IDS grant NAG5-12458. MT
acknowledges support from the University of Colorado Astrophysics
Theory Program through grants NASA ATP NNX07AG77G and NSF AST
0707474. We thank Britton Smith, Mike Santos and Ravi Sheth for useful
suggestions and discussions and the referee for a thorough and
constructive report.


\clearpage

\clearpage

\clearpage

\begin{table}
\begin{center}
\caption{Main Parameters for Star Formation and Feedback \label{tab:params}}
\begin{tabular}{lcccc}
\tableline\tableline
Model & Halo MF &  $\epsilon_{PopIII}$ & $\epsilon_{PopII}$ & $f_{esc}$ \\
\small{Standard} & ST99 &  1/halo & $5 \times 10^{-3}$ & $0.5$ \\
\small{LowEscape} & ST99 & 1/halo & $5 \times 10^{-3}$ & $0.05$ \\
\small{NoEscape} & ST99 &  1/halo & $5 \times 10^{-3}$ & $0.0$ \\
\small{Ext\_J21} & ST99 &  1/halo & $5 \times 10^{-3}$ & $0.5$ \\
\small{PS} & PS76 &       1/halo & $5 \times 10^{-3}$ & $0.5$ \\ 
\small{MultiPopIII} & ST99 & $5 \times 10^{-3}$  & $5 \times 10^{-3}$ & $0.5$ \\
\small{MultiPopIII\_high\_eff} & ST99 & $5 \times 10^{-2}$&  $5 \times 10^{-2}$ & $0.5$ \\
\small{MultiPopIII\_Ly-$\alpha$} & ST99 & $H_2$: 1/halo; $T \geq 10^4 K$: $5 \times 10^{-3}$  & $5 \times 10^{-3}$ & $0.5$ \\
\tableline
\end{tabular}
\tablecomments{Summary of the parameters that we modify from run to
run in our star formation model.  The first column identifies the
model, the second reports the halo mass function used (ST99: Sheth \&
Tormen, PS76: Press \& Schechter). The third and the fourth columns
contains the star formation efficiency for Population III and for
Population II stars. The last column the escape fraction. Model
Ext\_J21 includes an external radiative field (see
Eq.~\ref{eq:ext_LW}).}
\end{center}
\end{table}


\end{document}